%% file: siam-cse-report-aug-2017.tex
\documentclass[10pt]{article}

\usepackage[margin=1.10in]{geometry}
\usepackage{amssymb}
\usepackage{pdfpages}
\usepackage{graphicx}           
\usepackage{amsmath}            
\usepackage{float}              
\usepackage{verbatim}           
\usepackage{amssymb}            
\usepackage{url}                
\usepackage{times}              
\usepackage{color}
\usepackage{fancyvrb}
\usepackage{subfigure}
\usepackage{algorithm}
\usepackage{algorithmicx}
\usepackage{algpseudocode}
\usepackage{mathabx}
\usepackage{datetime}
\usepackage[export]{adjustbox}  
\usepackage{tcolorbox} 
\usepackage{capt-of}
\usepackage{wrapfig}
\usepackage[framemethod=tikz]{mdframed}
\definecolor{mycolor}{rgb}{0.122, 0.435, 0.698}
\newmdenv[userdefinedwidth=.1\linewidth,innerlinewidth=0.5pt, roundcorner=4pt,linecolor=mycolor,innerleftmargin=6pt,
innerrightmargin=6pt,innertopmargin=6pt,innerbottommargin=6pt]{mybox}

\newcommand{\team}[1] {}
\newcommand{\pagebudget}[1] {}

\usepackage{doi}

\definecolor{linkblue}{RGB}{0,0,140}
\usepackage{hyperref}
\hypersetup{
     colorlinks=true,
     citecolor=linkblue,
     filecolor=black,
     linkcolor=linkblue,
     urlcolor=linkblue
}
\usepackage{fancyhdr}
        {\begin{list}{\labelitemi}{
                \setlength{\topsep}{0pt}
                \setlength{\parskip}{0pt}
                \setlength{\itemsep}{0pt}
                \setlength{\parsep}{0pt}
                \setlength{\leftmargin}{23pt}
                \setlength{\labelwidth}{23pt}
        }}
        {\end{list}}

\renewcommand{\paragraph}[1]{\medskip \noindent {\em #1.~}}

\begin{document}

\pagestyle{empty}
\include{titlepage}

\pagenumbering{roman}
\pagestyle{fancy}
\rhead{}  
\lhead{\leftmark}
\cfoot{\thepage}

\input{dedication}
\mbox{} \clearpage
\input{abstract}

\mbox{} \clearpage

\newpage
\tableofcontents
\clearpage
\newpage

\pagenumbering{arabic}
\setcounter{page}{1}

\input{intro}


\input{research}

\input{education}

\input{conclusions}


\input{acknowledgments}

\addcontentsline{toc}{section}{Acknowledgments}

\newpage
\phantomsection
\addcontentsline{toc}{section}{References}

\bibliographystyle{siam}

\bibliography{cse,electron-beam-melting}

%
%

\end{document}

%% file: titlepage.tex
\begin{centering}
{\bf \Large Research and Education in Computational Science and Engineering}

\medskip
{\large August 2017}

\bigskip
  \includegraphics[width=0.50\textwidth]{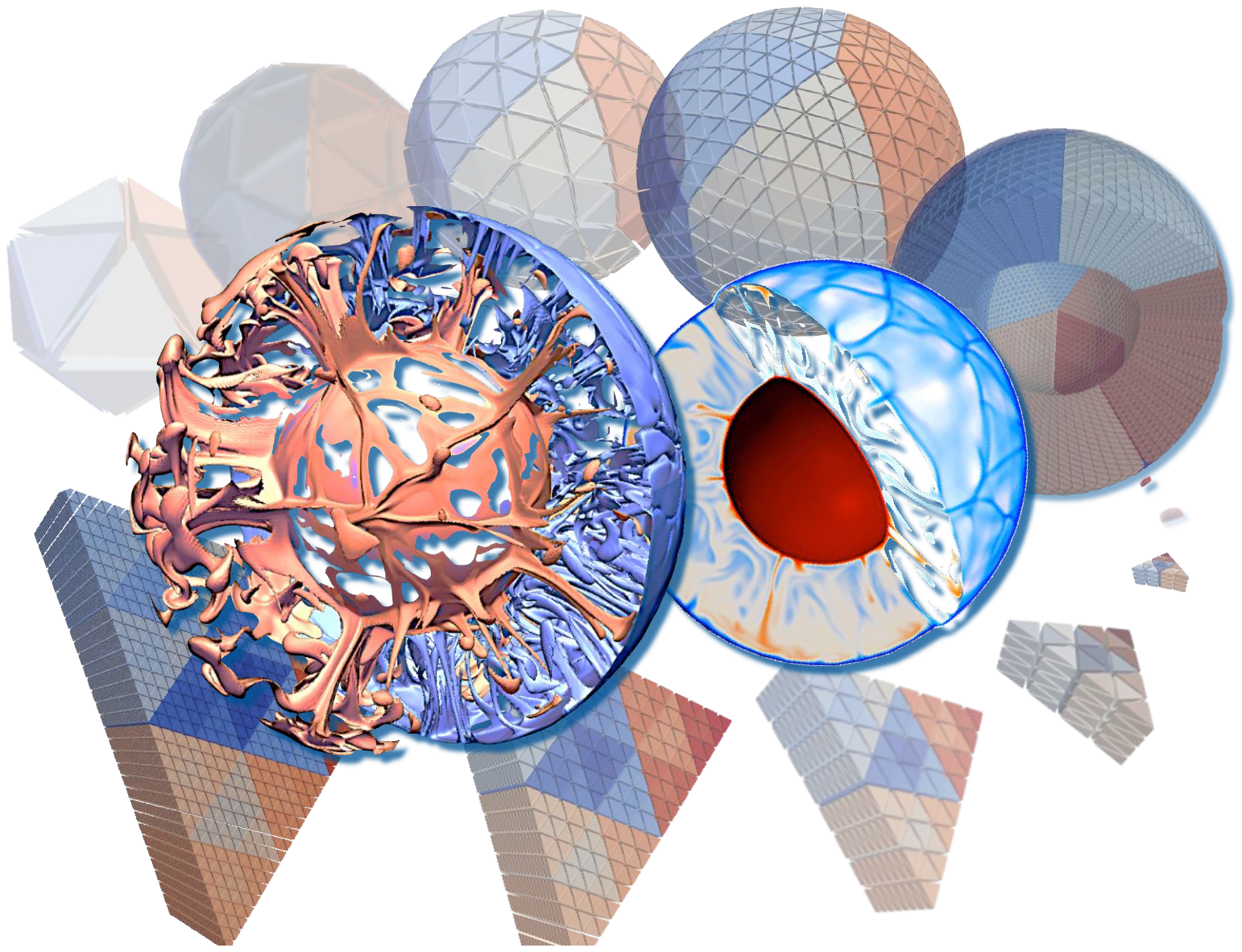}\\

\bigskip
{Report from a workshop sponsored by the Society for Industrial and Applied Mathematics (SIAM) and the European Exascale Software Initiative (EESI-2), August 4-6, 2014, Breckenridge, Colorado\\
}

\bigskip
{\bf Workshop Organizers:}\\
{\footnotesize 
Officers of the SIAM Activity Group on Computational Science and Engineering (SIAG/CSE), 2013-2014:\\

\medskip
Ulrich R\"{u}de, Universit\"{a}t Erlangen-N\"{u}rnberg, Chair\\
Karen Willcox, Massachusetts Institute of Technology, Vice Chair\\
Lois Curfman McInnes, Argonne National Laboratory, Program Director\\
Hans De Sterck, Monash University, Secretary\\
}

\medskip
{\bf Additional Contributors:}\\
{\footnotesize 
George Biros, University of Texas at Austin\\
Hans Bungartz, Technische Universit\"{a}t M\"{u}nchen\\
James Corones, Krell Institute\\
Evin Cramer, Boeing\\
James Crowley, SIAM\\
Omar Ghattas, University of Texas at Austin\\
Max Gunzburger, Florida State University\\
Michael Hanke, KTH Stockholm\\
Robert Harrison, Brookhaven National Laboratory and Stonybrook University\\
Michael Heroux, Sandia National Laboratories\\
Jan Hesthaven, \'{E}cole Polytechnique F\'{e}d\'{e}rale de Lausanne\\
Peter Jimack, University of Leeds\\
Chris Johnson, University of Utah\\
Kirk E. Jordan, IBM\\
David E. Keyes, KAUST\\
Rolf Krause, Universit\`{a}̀ della Svizzera Italiana, Lugano\\
Vipin Kumar, University of Minnesota\\
Stefan Mayer, MSC Software, Munich\\
Juan Meza, University of California, Merced\\
Knut Martin M{\o}rken, University of Oslo\\
J. Tinsley Oden, University of Texas at Austin\\
Linda Petzold, University of California, Santa Barbara\\
Padma Raghavan, Vanderbilt University\\
Suzanne M. Shontz, University of Kansas\\
Anne Trefethen, University of Oxford\\
Peter Turner, Clarkson University\\
Vladimir Voevodin, Moscow State University\\
Barbara Wohlmuth, Technische Universit\"{a}t M\"{u}nchen\\
Carol S. Woodward, Lawrence Livermore National Laboratory\\
}

\vspace{-0.25in}
\includegraphics[width=0.20\textwidth,padding*=0 0.22in 0 0]{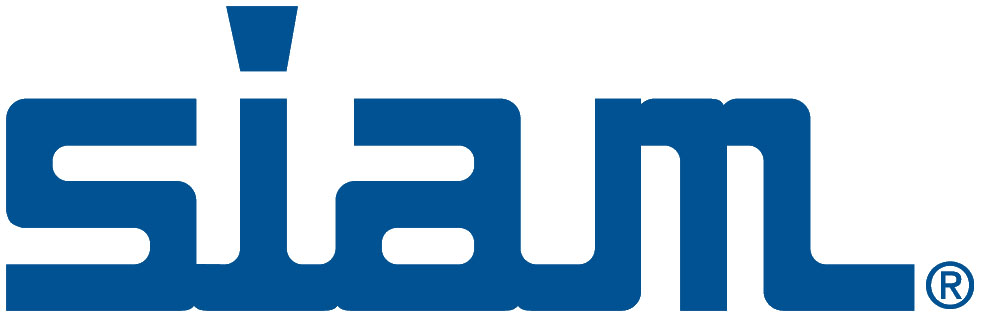}
\hspace{3.4in}
\includegraphics[width=0.20\textwidth]{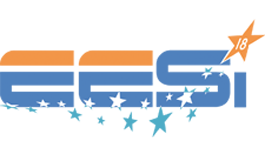}

\end{centering}

%% file: dedication.tex
$\,$
\vspace{3.0in}

\begin{center}
\noindent
This document is dedicated to the memory of Hans Petter Langtangen
(1962--2016),\\ a passionate scholar, teacher, and advocate of computational science and engineering.
\end{center}

%% file: abstract.tex
$\,$

\bigskip

\begin{center}
{\bf \Large Abstract}
\end{center}

\bigskip
\bigskip
\bigskip

This report presents challenges, opportunities and directions for computational science and engineering (CSE) research and education for the next decade. Over the past two decades the field of CSE has penetrated both basic and applied research in academia, industry, and laboratories to advance discovery, optimize systems, support decision-makers, and educate the scientific and engineering workforce. Informed by centuries of theory and experiment, CSE performs computational experiments to answer questions that neither theory nor experiment alone is equipped to answer. CSE provides scientists and engineers with algorithmic inventions and software systems that transcend disciplines and scales. CSE brings the power of parallelism to bear on troves of data. Mathematics-based advanced computing has become a prevalent means of discovery and innovation in essentially all areas of science, engineering, technology, and society; and the CSE community is at the core of this transformation.
However, a combination of disruptive developments---including the architectural complexity of extreme-scale computing,
the data revolution and increased attention to data-driven discovery, and the specialization required to follow the applications to new frontiers---is redefining the scope and reach of the CSE endeavor. With these many current and expanding opportunities for the CSE field, there is a growing demand for CSE graduates and a need to expand CSE educational offerings. This need includes CSE programs at both the undergraduate and graduate levels, as well as continuing education and professional development programs, exploiting the synergy between computational science and data science. Yet, as institutions consider new and evolving educational programs, it is essential to consider the broader research challenges and opportunities that provide the context for CSE education and workforce development.

\vspace{2.in}
\noindent
{\small
{\bf Cover Image:}
The cover image illustrates Earth mantle convection, the planetwide creeping flow of the Earth's mantle on time scales
of millions of years. The computational mesh is generated from an icosahedron
that is hierarchically refined 12 times to reach a global resolution of
one kilometer, resulting in a finite-element mesh with more than one trillion ($10^{12}$)
degrees of freedom.
Petascale-class machines and highly efficient, parallel multigrid methods
are required in order to solve the resulting equations in a timestepping procedure.
The image first appeared in the 2014 DFG mathematics calendar
(\url{http://www.dfg.de/sites/flipbook/dfg\_kalender\_2014/})
and is an outcome of the DFG project TerraNeo (PIs H.-P. Bunge, U. R\"ude, and B. Wohlmuth)
in the Priority Programme 1648
{\em Software for Exascale Computing}.}

\vspace{.4in}
\noindent
{\small
{\bf Please cite this document as follows:}
{Research and Education in Computational Science and Engineering},
Ulrich R\"{u}de, Karen Willcox, Lois Curfman McInnes, Hans {De Sterck},
George Biros, Hans Bungartz, James Corones, Evin Cramer,
James Crowley, Omar Ghattas, Max Gunzburger, Michael Hanke,
Robert Harrison, Michael Heroux, Jan Hesthaven, Peter Jimack,
Chris Johnson, Kirk E. Jordan, David E. Keyes, Rolf Krause, Vipin Kumar,
Stefan Mayer, Juan Meza, Knut Martin M{\o}rken, J.~Tinsley Oden,
Linda Petzold, Padma Raghavan, Suzanne M. Shontz, Anne Trefethen,
Peter Turner, Vladimir Voevodin, Barbara Wohlmuth, and Carol S. Woodward,
August 2017,
\href{http://arxiv.org/abs/1610.02608}{arXiv:1610.02608 [cs.CE]},
to appear in {\em SIAM Review}.
}

%% file: intro.tex
\section{CSE: Driving Scientific and Engineering Progress}
\label{sec:intro}
\pagebudget{5}
\subsection{Definition of CSE}
\begin{figure}[hb]
\centering
 \includegraphics[width=0.38\textwidth]{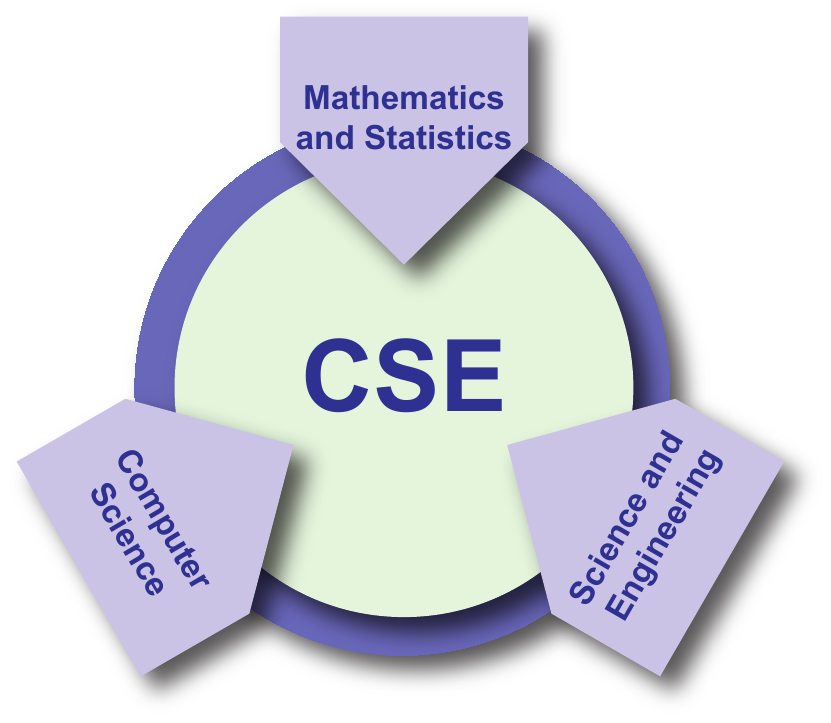}
\caption{CSE at the intersection of mathematics and statistics, computer science, and core disciplines from the sciences and engineering. This combination gives rise to a new field whose character is different from its original constituents.
\label{Fig:CSE-diagram}
}
\end{figure}

Computational science and engineering (CSE) is a multidisciplinary
field lying at the intersection of
mathematics and statistics, computer science, and core disciplines of
science and engineering (Figure~\ref{Fig:CSE-diagram}).
While CSE builds on these disciplinary areas, its focus is on the
integration of knowledge and methodologies from all of them and the
development of new ideas at their interfaces. As such, CSE is a field
in its own right, distinct from any of the core disciplines.
CSE is devoted to the development and use of computational methods for
scientific discovery in all branches of the sciences, for the
advancement of innovation in engineering and technology, and for the
support of decision-making across a spectrum of societally important
application areas.
CSE is a broad and vitally
important field encompassing methods of high-performance computing
(HPC) and playing a central role in the data revolution.

While CSE is rooted in the mathematical and statistical sciences,
computer science, the physical sciences, and engineering, today it
increasingly pursues its own unique research agenda.  CSE is now widely
recognized as an essential cornerstone that drives scientific and
technological progress in conjunction with theory and experiment.
Over the past two decades CSE has grown beyond its classical roots
in mathematics and the physical sciences
and has started to revolutionize
the life sciences and medicine.
In the 21st century its pivotal
role continues to expand to broader areas that now include the social
sciences, humanities, business, finance, and government policy.

\subsection{Goal of This Document}
The 2001 report on graduate education in CSE by the SIAM Working Group
on CSE Education \cite{education2001graduate} (see page \pageref{box:Petzold-report}) 
was instrumental in
setting directions for the then nascent CSE field. While its target
focus was CSE education, the report more broadly
emphasized the critical need to consider research and education together when contemplating future directions. Thus, recognizing that much has changed since the 2001 report,
the goal of this document is twofold: (1) examine and assess the rapidly expanding role of CSE in the 21st-century landscape of research and education and (2) discuss new directions for CSE research and education
in the coming decade.
We explore challenges and opportunities across CSE methods,
algorithms, and software, while examining the impact of disruptive
developments resulting from emerging extreme-scale computing systems,
data-driven discovery, and comprehensive broadening of the
application fields of CSE.
We discuss particular advances in CSE methods and algorithms,
the ubiquitous parallelism of all future computing, the sea change provoked by the data revolution and the synergy with data science,
the importance of software as a foundation for sustained CSE collaboration,
and the resulting challenges for CSE education
and workforce development.

\medskip
\begin{tcolorbox}[title=CSE Success Story: SIAM Working Group Inspires Community to Create CSE Education Programs]
\begin{wrapfigure}{R}{0.68\textwidth}
\vspace{-0.15in}
\includegraphics[width=0.68\textwidth]{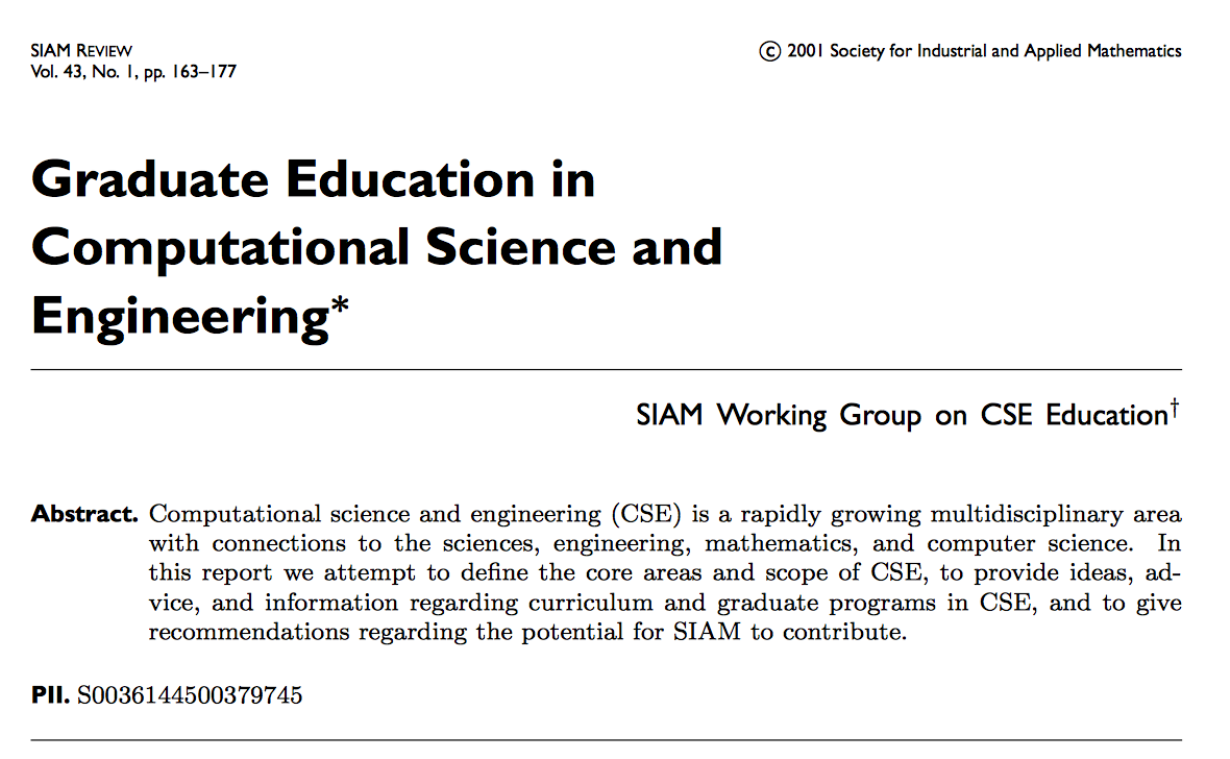}
\label{box:Petzold-report}
\end{wrapfigure}
The landmark 2001 report on ``Graduate Education in Computational Science and Engineering'' by L.~Petzold et al.~\cite{education2001graduate} played a critical role in helping define the then-nascent field of CSE. The report proposed a concrete definition of CSE's core areas and scope, and it laid out a vision for CSE graduate education. In doing so, it contributed a great deal to establishing CSE's identity, to identifying CSE as a priority interdisciplinary area for funding agencies, to expanding and strengthening the global offerings of CSE graduate education, and ultimately to creating the current generation of early-career CSE researchers.

\smallskip
Much of the 2001 report remains relevant today; yet much has changed. Sixteen years later, there is a sustained significant demand for a workforce versed in mathematics-based computational modeling and simulation, as well as a high demand for graduates with the interdisciplinary expertise needed to develop and/or utilize computational techniques and methods in many fields across science, engineering, business, and society. These demands necessitate that we continue to strengthen existing programs as well as leverage new opportunities to create innovative programs.
\end{tcolorbox}

\subsection{Importance of CSE } 

The impact of CSE on our society
has been so enormous---and the role of modeling and simulation so
ubiquitous---that it is nearly impossible to measure CSE's impact and
too easy to take it for granted. It is hard to imagine the design or
control of a system or process that has not been thoroughly
transformed by predictive modeling and simulation. Advances in CSE
have led to more efficient aircraft, safer cars, higher-density
transistors, more compact electronic devices, more powerful chemical
and biological process systems, cleaner power plants,
higher-resolution medical imaging devices, and more
accurate geophysical exploration technologies---to name just a
few.
A rich variety of fundamental advances have been enabled by CSE in areas such as
astrophysics,
biology,
climate modeling,
fusion-energy science,
hazard analysis,
human sciences and policy,
management of greenhouse gases,
materials science,
nuclear energy,
particle accelerator design, and
virtual product design
\cite{scales03,pitac05,OdenEtAl06,Brown08,Breakthroughs08,GlotzerKimEtAl2009,BrownMessina10,oden2011grand,KeyesMcInnesWoodwardEtAl13,Langtangen2015}.

\smallskip
\noindent
{\bf CSE as a complement to theory and experiment.}
CSE closes the centuries-old gap between theory and experiment by
providing technology that converts theoretical models into
predictive simulations. It creates a systematic method to
integrate experimental data with algorithmic models.
CSE has become the essential driver for progress in
science when classical experiments or conventional theory
reach their limits,
and in applications where experimental approaches are too
costly, slow, dangerous, or impossible. Examples include automobile
crash tests, nuclear test explosions, emergency flight maneuvers,
and operator emergency response training.
Experiments in fundamental
science may be impossible when the systems under study span
microscopic or macroscopic scales in space or time that are beyond
reach. Although traditional theoretical analysis would not suffer
from these limitations,
theory alone is insufficient to create predictive capabilities.  For
example, while the well-established mathematical models for fluid
dynamics provide an accurate theoretical description of the
atmosphere, the equations elude analytical solutions for problems of
interest because of their nonlinearity. When combined with the power
of numerical simulation and techniques to assimilate vast amounts of measured data,
these mathematical models become useful for complex problems such as predicting tomorrow's weather or
designing more energy-efficient aircraft wings.
Another example is the use of simulation models to
conduct systematic virtual experiments of exploding supernovae: CSE
technology serves as a virtual telescope reaching farther than any
real telescope, expanding human reach into outer space. And
computational techniques can equally well serve as a virtual
microscope, being used to understand quantum phenomena at scales so
small that no physical microscope could resolve them.

\smallskip
\noindent
{\bf CSE and the data revolution.}
The emergence and growing importance of massive data sets in many
areas of science, technology, and society, in conjunction with the
availability of ever-increasing parallel computing power, are
transforming the world. Data-driven approaches
enable novel ways of scientific discovery.
Using massive amounts of data and mathematical techniques to assimilate the
data in computational models offers new ways of quantifying
uncertainties in science and engineering and thus helps
make CSE truly predictive.
At the same time, relying on new forms of massive data, we can now
use the scientific approach of quantitative, evidence-based analysis
to drive progress in many areas of society
where qualitative forms of analysis, understanding, and decision-making
were the norm until recently.
Here the CSE paradigm contributes as a keystone technology to the
data revolution, in synergy with data science.

\begin{figure}[t]
\centering
\includegraphics[width=0.72\textwidth]{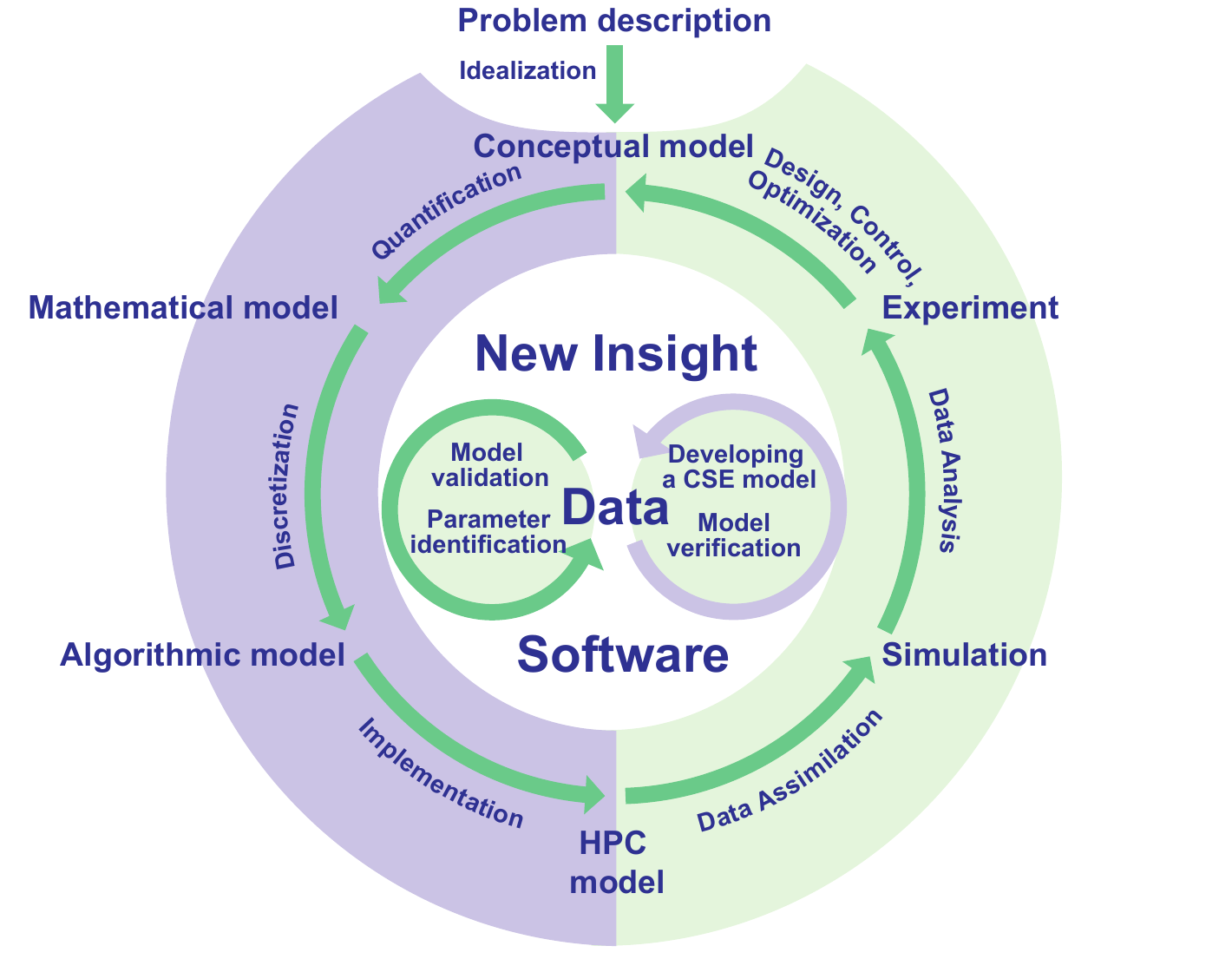}
\caption{CSE cycle---from physical problem to model and algorithms
  to efficient implementation in simulation software with verification
  and validation driven by data---leading to new insight in science and engineering.
\label{Fig:CSE-pipeline}}
\end{figure}

\smallskip
\noindent
{\bf CSE cycle.}
Many CSE problems can
be characterized by a {\em cycle} that includes mathematical
modeling techniques (based on physical or other principles),
simulation techniques (such as discretizations of equations and scalable
solvers),
and analysis techniques (data mining, data management,
and visualization, as well as the analysis of error, sensitivity,
stability, and uncertainty)---all encapsulated in high-performance scientific
software.
The CSE cycle is more than a sequential pipeline since it is
connected through multiple feedbacks, as illustrated in
Figure~\ref{Fig:CSE-pipeline}.
Models are revised and updated with new data.
When they reach a sufficient
level of predictive fidelity, they can be used for design and control,
which are often posed formally as optimization problems.

\smallskip
\noindent
{\bf CSE success stories.}
Throughout this document, we highlight a few examples of CSE success stories
in call-out boxes to illustrate how combined advances in CSE
theory, analysis, algorithms, and software have made CSE technology
indispensable for applications throughout science and industry.

\medskip
\begin{tcolorbox}[title=CSE Success Story: Computational Medicine]
\begin{wrapfigure}{R}{0.61\textwidth}
\vspace{-0.15in}
\includegraphics[width=0.61\textwidth]{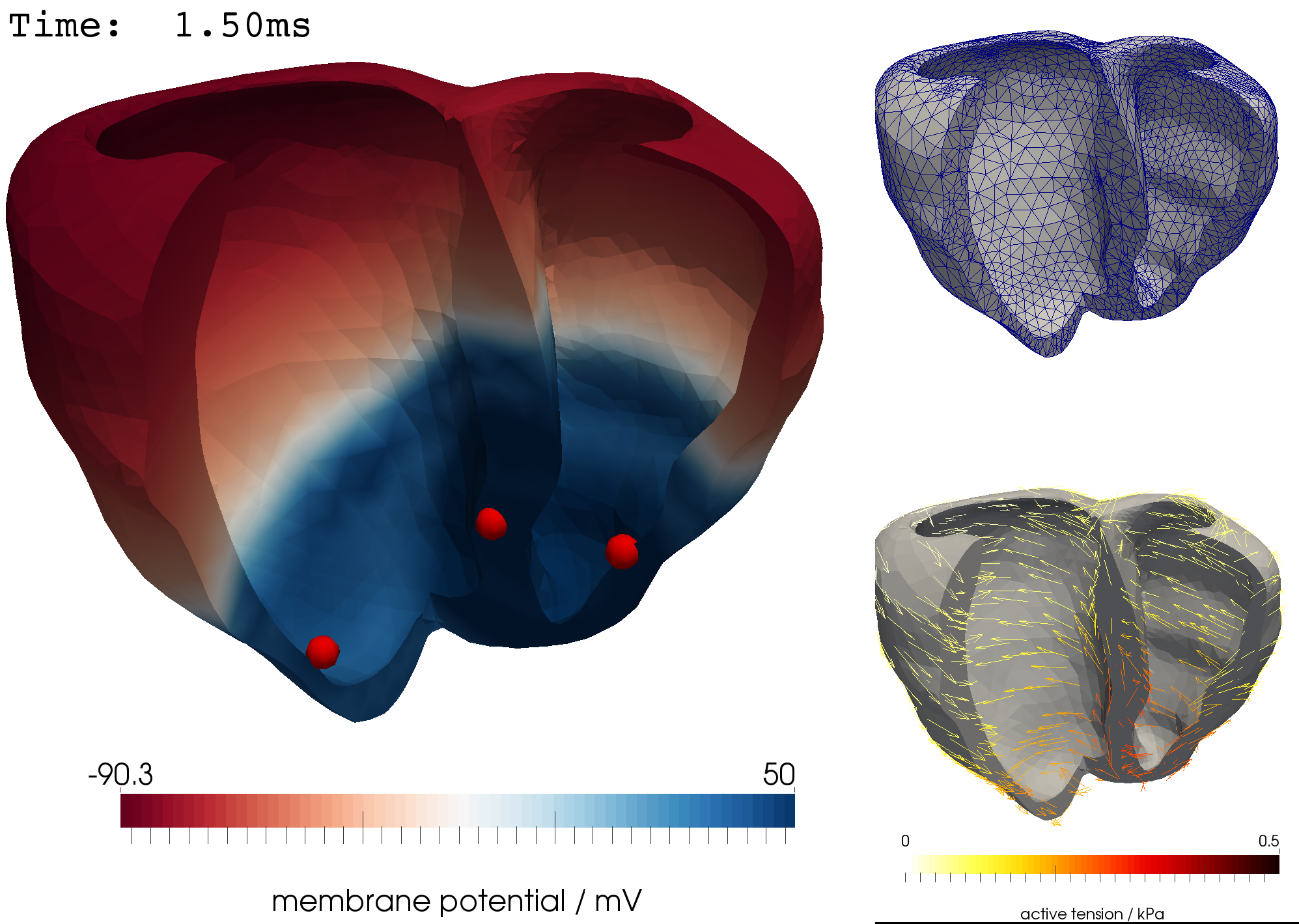}
\end{wrapfigure}
Computational medicine has always been at the frontier of CSE: the virtual design
and testing of new drugs and therapies accelerate medical progress and
reduce cost for development and treatment.
For example, CSE researchers have developed elaborate models of the electromechanical
activity of the human heart.\protect\footnotemark
~Such complex processes within the human body lead to elaborate multiscale models.
Cardiac function builds on a complicated interplay between different temporal
and spatial scales (i.e., body, organ, cellular, and molecular
levels), as well as different physical models (i.e., mechanics,
electrophysiology, fluid mechanics, and their interaction).
CSE advances in computational medicine are helping, for example, in placing electrodes
for pacemakers and studying diseases such as atrial fibrilation.
Opportunities abound for next-generation CSE advances:
The solution of inverse problems can help
identify suitable values for material parameters, for example, to detect scars
or infarctions. Using uncertainty quantification, researchers can estimate the
influence of varying these parameters or varying geometry.
\end{tcolorbox}

\footnotetext{Parallel and adaptive simulation method described in T. Dickopf, T. Krause, R. Kraus, and M. Potse, SIAM J Sci Comput 36(2), C163-C189, 2014.}

\subsection{Challenges and Opportunities for the Next Decade}

While the past several decades have witnessed tremendous progress in
the development of CSE methods and their application within a broad
spectrum of science and engineering problems, a number of challenges
and opportunities are arising that define important research directions
for CSE in the coming decade.

In science and engineering simulations, large differences in temporal and
spatial scales must be
resolved together with handling uncertainty in parameters and
data, and often different models must be coupled together
to become complex multiphysics simulations.
This integration is necessary in order to tackle \textbf{applications in a multitude of new fields} such as
the biomedical sciences. High-fidelity predictive simulations require
feedback loops that involve inverse problems, data assimilation,
and optimal design and control.
Algorithmic advances in these areas
are at the core of CSE research; and in order to deal with the requirements of
ever more complex science and engineering applications, new
\textbf{fundamental mathematical and algorithmic developments} are required.

Several recent disruptive developments yield the promise of further fundamental
progress if new obstacles can be overcome.
Since single-processor clock speeds have stagnated,
any further increase in computational power must result from a
further increase in parallelism.
New mathematical and computer science techniques need to be explored that can
guide development of modern algorithms that are effective in the
new era of \textbf{ubiquitous parallelism} and extreme-scale computing.
In addition, the sea change provoked by the data revolution requires new methods for
\textbf{data-driven scientific discovery} and new algorithms for \textbf{data analytics}
that are effective at very large scale, as part of the comprehensive
broadening of the application fields of CSE to almost every field of
science, technology, and society, in synergy with data science.
Moreover, software itself is now broadly recognized as a key crosscutting technology that connects advances
in mathematics, computer science, and domain-specific science and engineering
to achieve robust and efficient simulations on advanced computing systems.
In order to deliver the comprehensive CSE promise, the role of
{\bf CSE software ecosystems} must be
redefined---encapsulating advances in algorithms, methods, and
implementations and thereby providing critical instruments to enable
scientific progress.

These exciting research challenges and opportunities will be
elaborated on in Section \ref{sec:research} of this document.

\subsection{The Broad CSE Community}
\label{sec:cse-community}

The past two decades have seen tremendous growth in the CSE community,
including a dramatic increase in both the size and breadth of intellectual
perspectives and interests. The growth in community size can be seen,
for example, through the membership of the SIAM Activity Group on CSE,
which has steadily increased from approximately 1,000 members in 2005 to more than 2,500
in 2017.
The biennial SIAM CSE Conference~\cite{siam-cse-conference-series}
is now SIAM's largest conference, with growth
from about 400 attendees in 2000 to over 1,700 attendees in 2017.
The increased breadth
of the community is evidenced in many ways: by the diversity of
minisymposium topics at SIAM CSE conferences; through a new broader
structure for SIAM's {\em Journal on Scientific Computing}, including a new
journal section that focuses on computational methods in specific
problems across science and engineering; and by the sharply increased
use of CSE approaches in industry~\cite{oden2011grand,HEC-IWG2013}.

\begin{figure}[htb]
\centering
\includegraphics[width=0.65\textwidth]{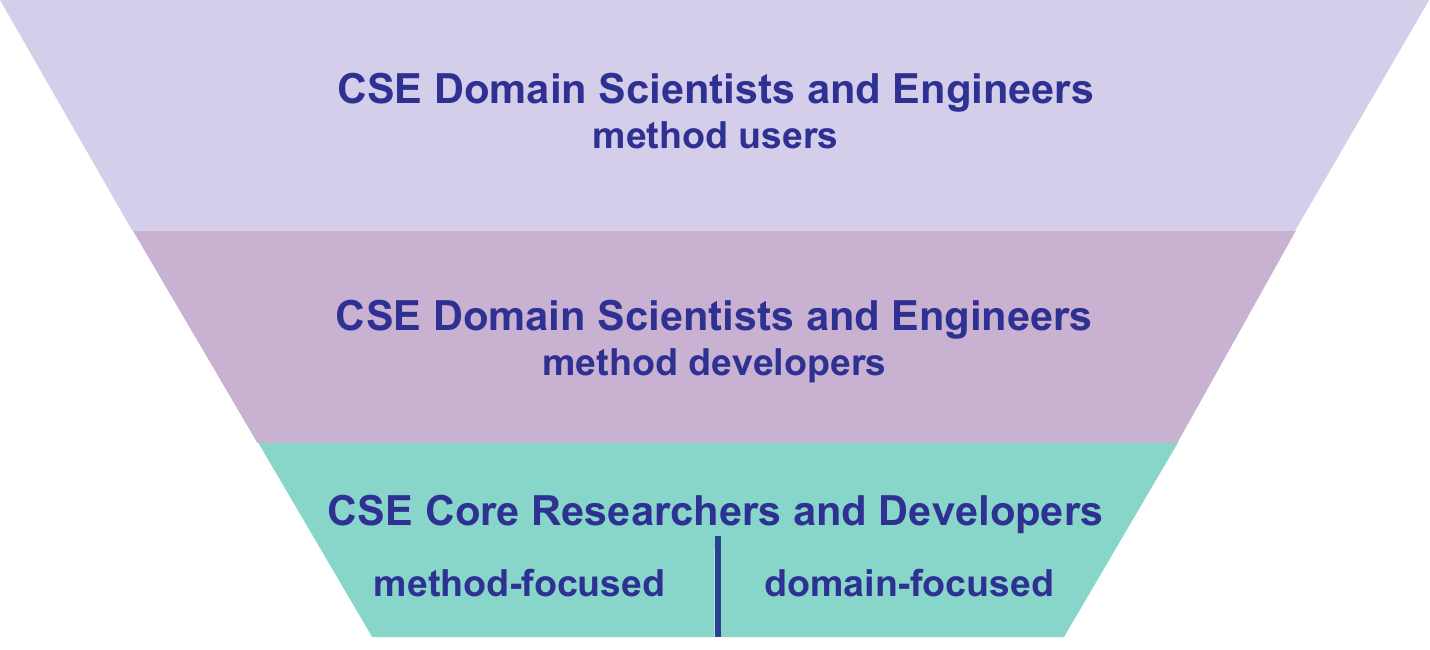}
\caption{One view of the different aspects of the broad CSE community. The part of the CSE community that focuses on developing new methods and algorithms is labeled {\em CSE Core Researchers and Developers}. This group may be driven by generally applicable methods or by methods developed for a specific application domain. {\em CSE Domain Scientists and Engineers} focus their work primarily in their scientific or engineering domain and make extensive use of CSE methods in their research or development work.
}
\label{Fig:CSE-community}
\end{figure}

As we envision the future of CSE, and in particular as we consider
educational programs, we must keep in mind that such a large and diverse
intellectual community has a correspondingly broad set of
needs. Figure~\ref{Fig:CSE-community} presents one way to view the
different aspects of the broad CSE community: (1) \textit{CSE Core
  Researchers and Developers}---those engaged in the conception,
analysis, development, and testing of CSE algorithms and software and
(2) \textit{CSE Domain Scientists and Engineers}---those primarily engaged
in developing and exploiting CSE methods for progress in
particular science and engineering campaigns.
The latter community can usefully be further
categorized into those who interact with the core technologies at a
developer level within their own applications, creating their own
implementations and contributing to methodological/algorithmic
improvements, and those who use state-of-the-art CSE technologies as
products, combining them with their expert knowledge of an application
area to push the boundaries of a particular application.  Within the
\textit{CSE Core Researchers and Developers} group in
Figure~\ref{Fig:CSE-community}, we further identify two groups: those
focused on broadly applicable methods and algorithms and those focused
on methods and algorithms motivated by a specific domain of
application. This distinction is a useful way to
cast differences in desired outcomes for different types of CSE
educational programs
as they will be discussed in Section
\ref{sec:education}.
As with any such categorization, the dividing lines
in Figure~\ref{Fig:CSE-community} are fuzzy, and in fact any single
researcher might span multiple categories.

\subsection{Organization of This Document}

The remainder of this document is organized as follows.
Section~\ref{sec:research} presents challenges and opportunities in
CSE research, organized into four main areas. First we discuss
key advances in core CSE methods and algorithms, and the
ever-increasing parallelism in computing hardware culminating
in the drive toward extreme-scale applications.
Next we describe how the ongoing data revolution offers tremendous
opportunities for breakthrough advances in science and engineering by
exploiting new techniques and approaches in synergy with data science,
and we discuss the challenges in advancing CSE software
given its key role as a crosscutting CSE technology.
Section~\ref{sec:education} discusses how major changes in
the CSE landscape are affecting the needs and goals of CSE education
and workforce development.  Section~\ref{sec:conclusions} summarizes findings and formulates
recommendations for CSE over the next
decade.

%% file: research.tex
\section{Challenges and Opportunities in CSE Research}
\label{sec:research}

The field of CSE faces important challenges and opportunities for the next decade, following disruptive developments in extreme-scale computing and ubiquitous parallelism, the emergence of big data and data-driven discovery, and a comprehensive broadening of the application areas of CSE. This section highlights important emerging developments in CSE methods and algorithms, in HPC, in data-driven CSE, and in software.

\input{advances-cse-core}

\input{hpc-cse}

\input{data-cse}

\input{software-cse}

\input{predictive-cse}

%% file: advances-cse-core.tex
\subsection{Advances in CSE through Mathematical Methods and Algorithms}
\label{sec:core-cse}

Algorithms (e.g., see \cite{Higham2015}) occupy a central role in CSE.
They all transform inputs to outputs,
but they may differ in their generality, in their robustness and stability, and in their complexity---that is, in the way their costs in operations, memory, and data motion scale with the size of the input.
Mathematical theories and methods are of fundamental importance for the algorithms
developed and employed in CSE.
The types of algorithms employed in CSE are diverse.
They include geometric modeling, mesh generation and refinement,
discretization, partitioning, load balancing,
solution of ordinary differential equations (ODEs) and differential algebraic equations,
solution of partial differential equations (PDEs),
solution of linear and nonlinear systems, eigenvalue computations, sensitivity analysis,  error estimation and adaptivity,
solution of integral equations,
surrogate and reduced modeling, random number generation,
upscaling and downscaling between models, multiphysics coupling,
uncertainty quantification, numerical optimization, parameter identification, inverse problems,
graph algorithms, discrete and combinatorial algorithms, graphical models,
data compression, data mining, data visualization, and data analytics.

\subsubsection{Impact of Algorithms in CSE}
Compelling CSE success stories stem from breakthroughs in applied mathematics and computer science
that have dramatically advanced simulation capabilities through better algorithms, as
encapsulated in robust and reliable software.
The growing importance of CSE in increasingly many application areas has paralleled the exponential growth in computing power according to ``Moore's law''---the observation that over the past five decades the density of transistors on a
chip has doubled approximately every 18 months as a result of advances in lithography allowing miniaturization.
Less appreciated but crucial for the success of
CSE is the progress in algorithms in this time span.
The advances in computing power have been matched or even exceeded by
equivalent advances 
of the mathematics-based computational algorithms that lie at the heart of CSE.
Indeed, the development of efficient new algorithms 
has been crucial to the effective use of advanced computing
capabilities.
And as the pace of advancement in Moore's law
slows,\footnote{as documented in the TOP 500 list, \url{https://www.top500.org}} advances in algorithms and software will become even more important.
Single-processor clock speeds have stagnated, and further increase in computational power must come from increases in parallelism. CSE now faces the challenge of developing efficient methods and implementations in the context of ubiquitous parallelism (as discussed in Section \ref{sec:hpc-cse}).

\begin{figure}[h!]
\centering
\includegraphics[width=0.48\textwidth]{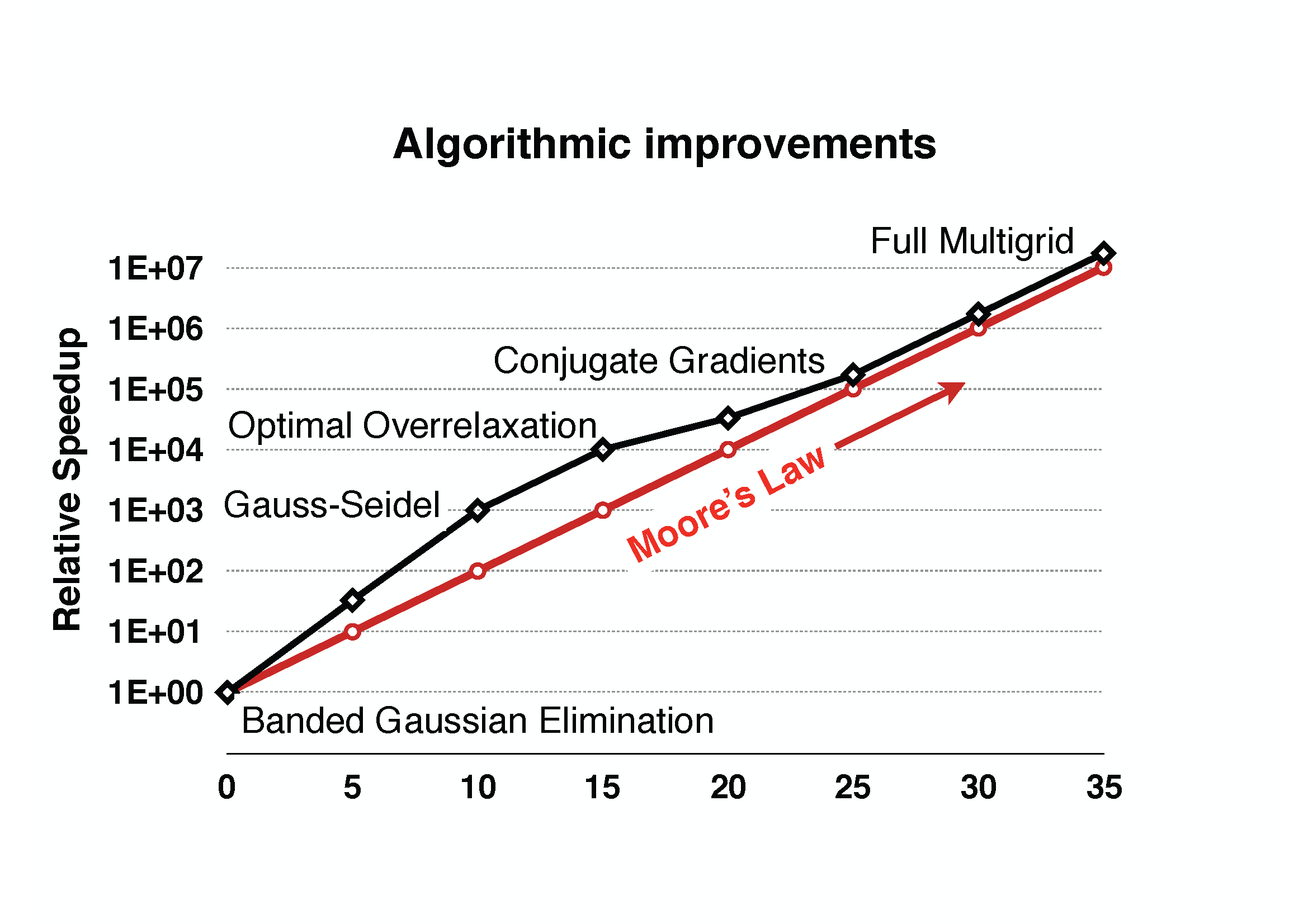}
\includegraphics[width=0.48\textwidth]{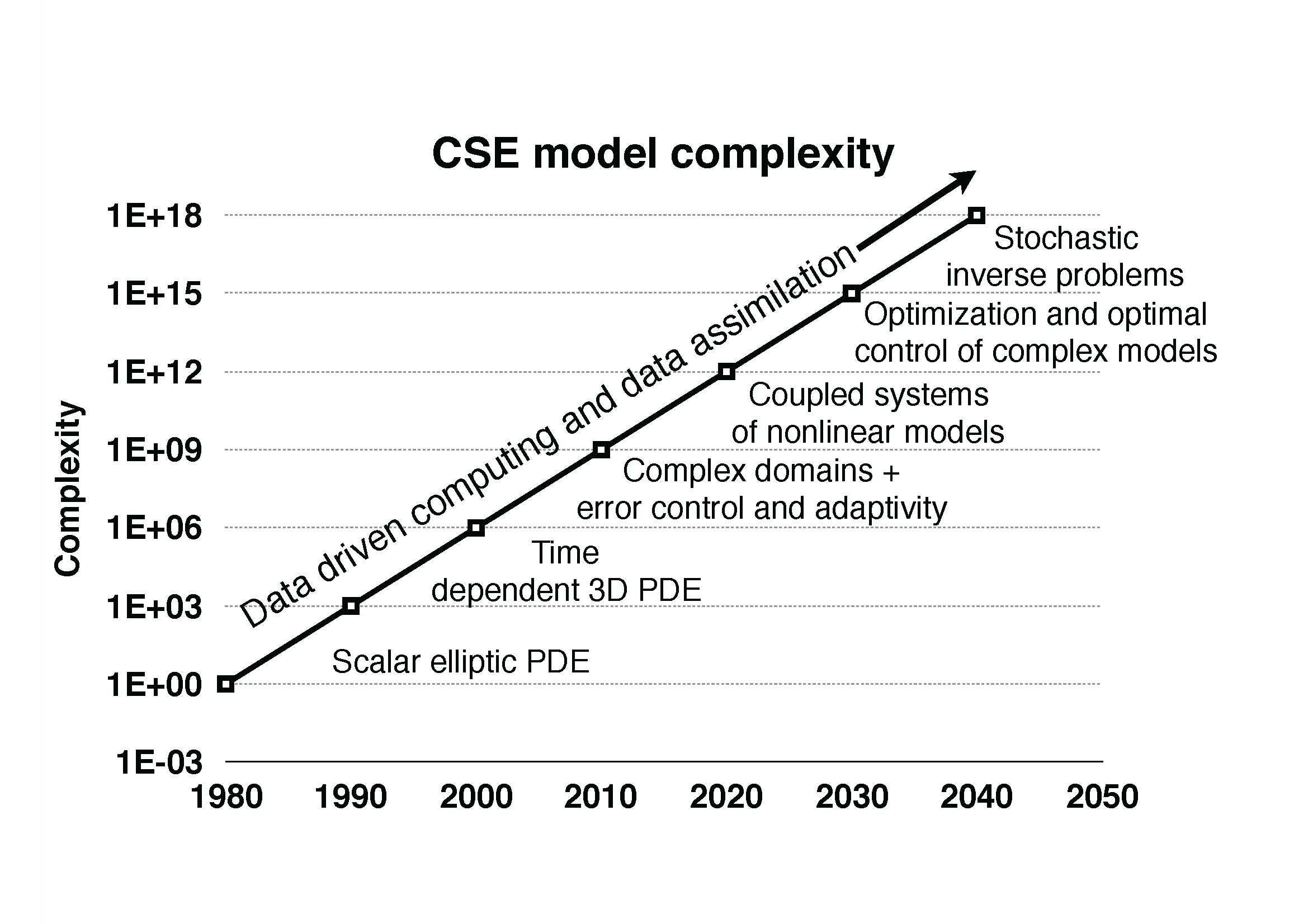}
\caption{\label{Fig:Algorithmic-Moore}
{\em Left:} Moore's law for algorithms to solve the 3D Poisson equation (black) plotted with Moore's law for transistor density (red), each showing 24 doublings (factor of approximately 16 million) in performance over an equivalent period.  For algorithms, the factor can be made arbitrarily large by increasing the problem size $N=n^3$. Here $n=64$, which is currently a modest resolution even on a single processor. {\em Right:} Increase in CSE model complexity and approximate computational cost over time, where the y-axis indicates a qualitative notion of complexity in the combination of models, algorithms, and data structures.  Simulations have advanced from modestly sized forward simulations in simple geometries to incorporate complex domains, adaptivity, and feedback loops.  The stage is set for new frontiers of work on advanced coupling, numerical optimization, stochastic models, and many other areas that will lead to truly predictive scientific simulations.}
\end{figure}

As problems scale in size and memory to address increasing needs for fidelity and resolution in grand-challenge simulations,
the computational complexity must scale as close to linearly in the problem size as possible.
Without this near-linear scaling, increasing memory and processing power in proportion---the way parallel computers are architected---the result will be computations that {\em slow down\/} in wall-clock time as they are scaled up.
In practice, such {\em optimal algorithms\/} are allowed to have a complexity
of $N(\log N)^p$, where $N$ is the problem size and $p$ is some small power, such as 1 or 2.
Figure~\ref{Fig:Algorithmic-Moore} illustrates the importance of algorithmic innovation since the beginnings of CSE.
We contrast here the importance of algorithmic research with technological progress in computers by using the historical example of linear solvers for elliptic PDEs.
Consider the problem of the
Poisson equation on a cubical domain, discretized into $n$ cells on a side, with a total problem size $N=n^3$.  The total memory occupied is $O(n^3)$, and the time to read in the problem or to write out the solution is also $O(n^3)$.
Based on the natural ordering, 
banded Gaussian elimination applied to this problem requires
$O(n^7)$  arithmetic operations.
Perhaps worse, the memory to store intermediate results bloats to $O(n^5)$---highly nonoptimal, so that if we initially fill up the memory with the largest problem that fits, it overflows.
Over a quarter of a century, from a paper by Von Neumann and Goldstine in 1947 to a paper by Brandt in 1974 describing optimal forms of multigrid, the complexity of both operations and storage was reduced, in a series of algorithmic breakthroughs, to an optimal $O(n^3)$ each. These advances are depicted graphically in a log-linear plot of effective speedup over time in the left-hand side of Figure 5.  During the same period, Moore's law accounted for approximately 24 doublings, or a factor of $2^{24} \approx 16$ million in arithmetic processing power per unit square centimeter of silicon, with approximately constant electrical power consumption.  This same factor of 
16 million was achieved
by mathematical research on algorithms
in the case that $n=2^6$ in the example above. 
For grids finer than $64 \times 64 \times 64$, as we routinely use today,
the progress of optimal algorithms overtakes the progress stemming from Moore's law, by an arbitrarily large factor.
Remarkable progress in multigrid has been made since
this graph
was first drawn.  Progress can be
enumerated along two directions: extension of optimality to problems
with challenging features not present in the original Poisson problem
and extension of optimal algorithms to the challenging environments of
distributed memory, shared memory, and hybrid parallelism, pushing
toward extreme scale.

Algorithmic advances of similar dramatic magnitudes across many areas
continue to be the core of CSE research.
These advances are often built on the development of new mathematical theories and methods.
Each decade since Moore stated his law in 1965,
computational mathematicians have produced new algorithms
that have revolutionized computing.
The impact of these algorithms in science and engineering,
together with the technological advances following Moore's law,
has led to the creation of CSE as a discipline and has enabled scientists to
tackle problems with increasing realism and complexity,
as shown in the right-hand side of Figure~\ref{Fig:Algorithmic-Moore}.

\medskip
\begin{tcolorbox}[title=CSE Success Story: Lightning-Fast Solvers for the Computer Animation Industry]
\begin{wrapfigure}{R}{0.7\textwidth}
\vspace{-0.15in}
\includegraphics[width=0.7\textwidth]{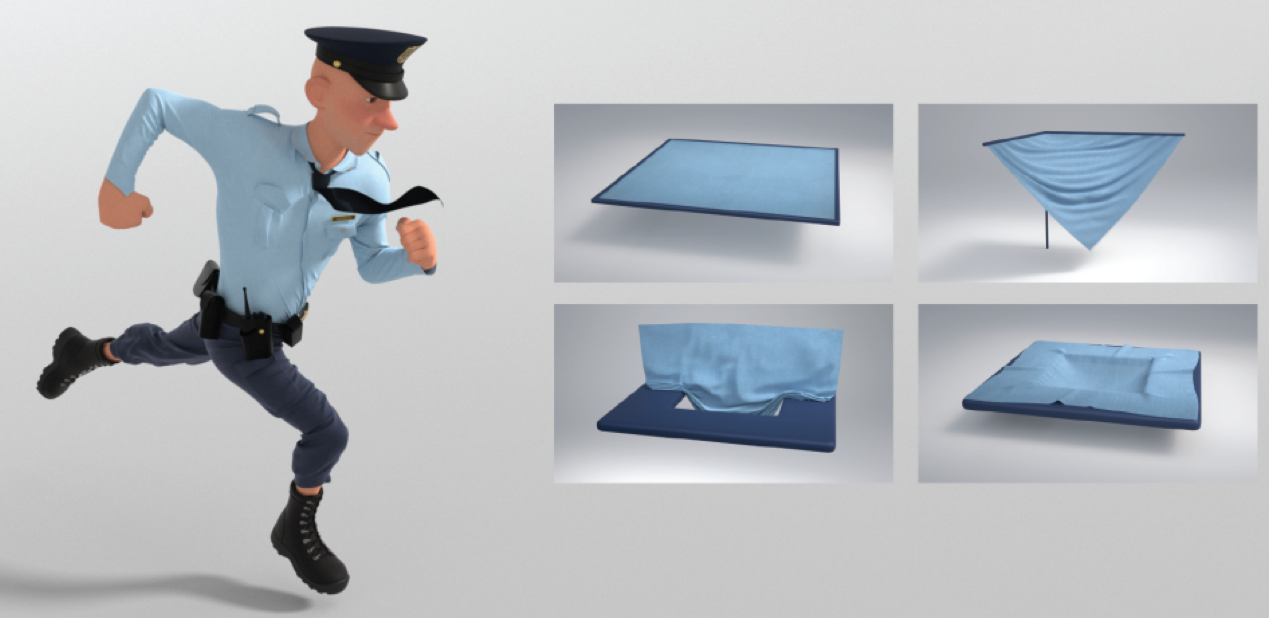}
\end{wrapfigure}
    CSE researchers have teamed up with computer animators at Walt Disney Animation Studios Research to dramatically improve the efficiency in linear system solvers that lie at the heart of many computer animation codes. Building on advanced multilevel methods originally developed for engineering simulations of elastic structures and electromagnetic systems,
    researchers showed that movie animations with cloth simulation on a fully dressed character discretized on an unstructured computational grid with 371,000 vertices could be accelerated by a factor of 6 to 8 over existing solution techniques.\protect\footnotemark ~These improvements in computational speed enable greater productivity and faster turnaround times for feature film production with realistic resolutions. Another application is real-time virtual try-on of garments in e-commerce.
\end{tcolorbox}

\footnotetext{See video at \url{https://youtu.be/_mkFBaqZULU} and paper by R. Tamstorf, T. Jones, and S. McCormick, ACM SIGGRAPH Asia 2015, at \url{https://www.disneyresearch.com/publication/smoothed-aggregation-multigrid/}.}

\subsubsection{Challenges and Opportunities in CSE Methods and Algorithms}

Without attempting to be exhaustive, we highlight here several areas of current interest where novel CSE algorithms have produced important advances. This often goes hand-in-hand with formulating new mathematical theories and their advancement. We also discuss challenges and opportunities for the next decade in these fields.

\input{solvers}

\input{uq}

\input{optimization}

\medskip
\begin{tcolorbox}[title=CSE Success Story: Bayesian Inversion for the
Antarctic Ice Sheet] 
\begin{wrapfigure}{R}{0.65\textwidth}
\vspace{-0.15in}
\includegraphics[width=0.65\textwidth]{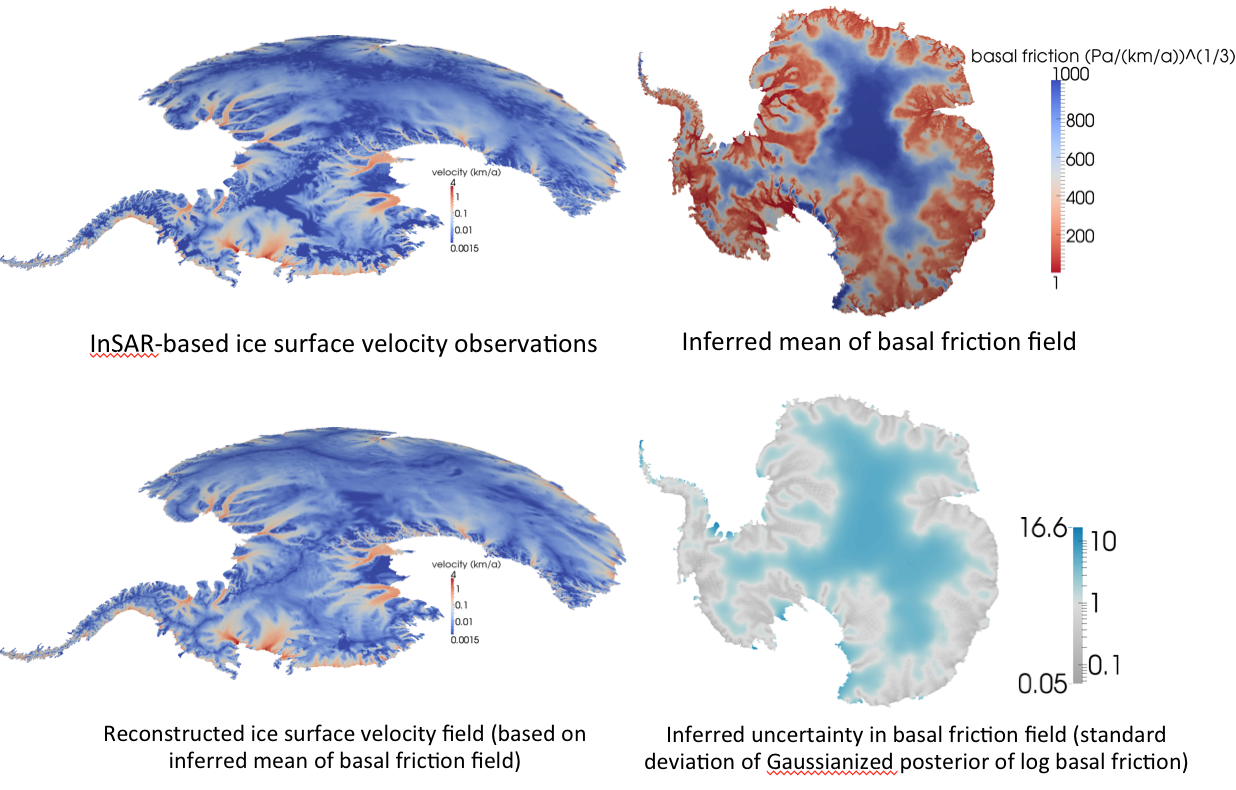}
\end{wrapfigure}
The question of how one infers unknown parameters characterizing a
given model of a physical system from observations of the outputs of
that model is fundamentally an inverse problem.  In order to address the
intrinsic ill-posedness of many inverse problems, regularization is
invoked to render the inverse solution unique. The Bayesian
formulation of the inverse problem seeks to infer all models, with
associated uncertainty, in the model class that are consistent with the data and any
prior knowledge. The figure illustrates the Bayesian solution of an
inverse problem to infer friction at the base of the Antarctic ice
sheet, from InSAR satellite observations of the surface ice flow
velocity and a non-Newtonian model of the flow. The upper left image
depicts the observed surface velocity field; the upper right image
shows the inferred basal friction field. The upper row thus
illustrates the classical regularization-based solution. The Bayesian
solution contains additional information about uncertainty in the
inverse solution, as illustrated in the lower right image of the
variance of the inferred basal friction. The lower left image
shows the predicted surface velocity field, using the inferred basal
friction. In order to overcome the prohibitive nature of large-scale Bayesian
inversion, the low-rank structure of the parameter-to-observable map
is exploited.
\protect\footnotemark
\end{tcolorbox}

\footnotetext{Figures from T. Isaac, N. Petra, G. Stadler, and O. Ghattas, Journal of Computational Physics 296, 348--368, 2015.}

\input{high-order-AMR}

\input{model-order-reduction}

\input{randomized-algorithms}

\input{multibody}

\input{multiphysics}

%% file: solvers.tex

\medskip
\noindent {\bf Linear, nonlinear, and timestepping solvers.} 
\label{sec:solvers}
As exemplified by the pursuit of optimality for algorithms for
elliptic PDEs described above, algebraic solvers receive extensive
attention because they represent the dominant computational cost and
dominant memory occupancy of many important CSE applications.  
In fact, five of the ``top ten algorithms of the twentieth century'' as
described by the guest editors of the January 2000 issue of {\em
Computing in Science \& Engineering\/} \cite{DongarraSullivan2000}
fall into the category of solvers.
Important types of problems requiring solvers that operate in
floating point include linear algebraic systems, nonlinear algebraic
systems, and eigensystems.
Timestepping algorithms for differential equations, after discretization,
also rely on algebraic solvers,
applying a forward operator for explicit techniques, an inverse operator for
implicit techniques, or a combination for implicit-explicit approaches.
Root-finding problems for nonlinear algebraic systems
are conventionally handled with linearization and Newton-type
iteration, focusing the main issue for progress in many CSE
applications on linear solvers.  Nevertheless, some recent algorithmic approaches
directly work with nonlinear algebraic systems.

A key approach in the pursuit of linearly scaling methods, for example for elliptic
PDE operators, is to employ a hierarchy of resolution scales. 
The most classical algorithm of this kind is the fast Fourier transformation (FFT) that is 
applicable in many important special cases.
Other prominent methods that rely on a hierarchy of scales are the multigrid, wavelet, multipole
and multilevel Monte Carlo methods, including their many extensions and descendants.
Another fruitful approach is to exploit known structures and properties arising from the underlying physics,
as, e.g., in physics-based preconditioning techniques and mimetic methods.
Geometric multigrid methods
iteratively form a hierarchy of problems derived from the original one on
a succession of coarser scales and remove on each scale the components
of the overall error most effectively represented on that scale.
Challenging features common in applications include inhomogeneity,
anisotropy, asymmetry, and indefiniteness.  
Geometric multigrid may lose its optimality 
on such challenging problems; however,
algebraic multigrid methods 
have revolutionized
many fields that were previously computationally intractable at finely
resolved scales.
Parallel preconditioners based on advanced domain decomposition approaches
also contribute to robustly scalable linear solvers.
Further progress in multigrid focuses, for example, on indefinite high-frequency
wave problems and efficient performance and resilience for 
extreme-scale architectures.

In such environments, data motion represents a higher cost than does arithmetic computation, 
and global synchronization makes algorithms
vulnerable to load and performance imbalance among millions or
billions of participating processors.
The fast multipole method possesses arithmetic intensity
(the ratio of flops to bytes moved) up to two orders of magnitude
greater, in some phases, than does the sparse matrix-vector multiply that
is the core of many other solvers.
In addition, fast multipole, being essentially a
hierarchically organized fast summation, is less vulnerable to
frequent synchronization. 
These properties have led fast multipole methods to
be considered as a replacement whenever an analytically evaluated
Green's function kernel is available.

Fast multipole is optimal because, for a given accuracy requirement,
it compresses interactions at a distance into coarse representations,
which can be translated at low cost and re-expanded locally,
relying on the ability to represent the interactions with operators of low
effective rank.
Many relevant operators 
have hierarchical low-rank structure even
when they do not admit a constructive Green's function expansion.
This structure enables replacement of the dominant coupling represented
in off-diagonal portions of a matrix by low-rank representations with
controllable loss of accuracy and major gains in storage, arithmetic
complexity, and communication complexity. Today, hierarchically
low-rank or ``rank-structured'' methods of linear algebra are
finding use in developing optimal solvers for an increasing class
of challenging problems.
Efficient approximate eigendecomposition methods are needed in order to generate low-rank
approximations to off-diagonal blocks in hierarchical matrices,
identifying the dominant subspace. Methods such as randomized singular value
decomposition are of specific interest.
While progress since \cite{education2001graduate} has been 
fruitful, the quest for 
powerful scalable solvers will likely
always be at the heart of CSE.

\medskip
\begin{tcolorbox}[title=CSE Success Story: Transforming the Petroleum Industry]
\begin{wrapfigure}{R}{0.7\textwidth}
\vspace{-0.15in}
\includegraphics[width=0.33\textwidth]{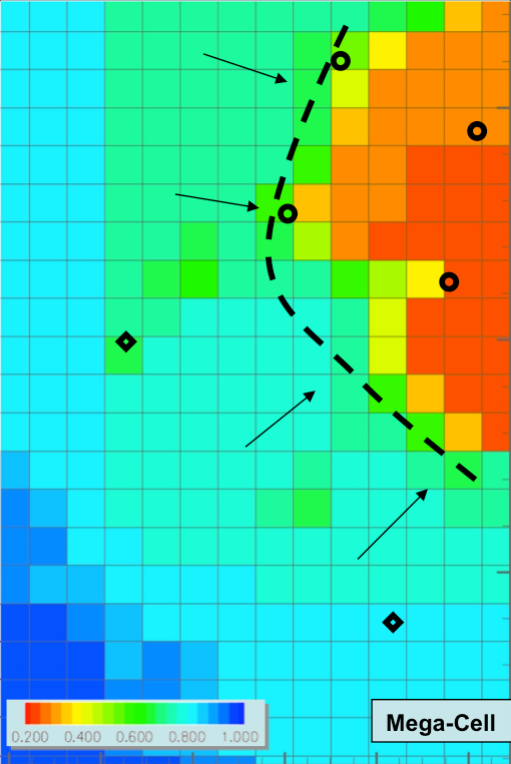}
\includegraphics[width=0.33\textwidth]{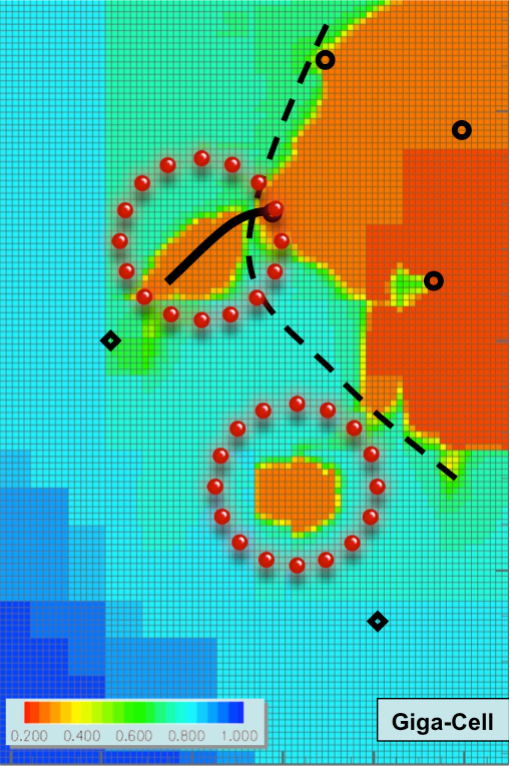}
\end{wrapfigure}
Few industries have been as transformed by CSE as petroleum, in which a decision to drill can commit \$100,000,000 or more. Reservoir imaging solves inverse problems (seismic and electromagnetic) to locate subsurface fluids in highly heterogeneous media and distinguish hydrocarbons from water.  Reservoir modeling simulates the flow of fluids between injection and production wells. Correctly predicting a pocket of oil left behind can justify an entire corporate simulation department.  Optimizing reservoir exploitation while reducing uncertainty requires simulating many forward scenarios. Oil companies have been behind the earliest campaigns to improve linear algebraic solvers and today operate some the world's most powerful computers. In the figure,\protect\footnotemark ~a reservoir is modeled with a coarse grid (left) and with a finer grid (right).  Diamonds are injection wells, and circles are production wells.  Unresolved on the coarse grid are two pockets of oil recoverable with horizontal drilling extensions.
\end{tcolorbox}

\footnotetext{Figure used by permission of Saudi Aramco.}

%% file: uq.tex

\newpage
\noindent {\bf Uncertainty quantification.}
\label{sec:uq}
Recent years have seen increasing recognition of the critical
role of uncertainty quantification (UQ) in all phases of CSE,
from inference to prediction to optimization to
decision-making. Just as the results of an experiment would not be meaningful
unless accompanied by measures of the uncertainty in the experimental
data, so too in CSE scientists need to know what confidence they can have in the predictions of
models. This issue is becoming urgent as CSE models become
increasingly for decision-making about critical technological and
societal systems.
As one indication of the explosion of interest in UQ in the past few years,
the recent 2016 SIAM UQ conference had more minisymposia than does the SIAM annual
meeting. Moreover, several U.S. federal agency-commissioned reports
focusing wholly or partially on status, opportunities, and challenges
in UQ have appeared in recent years \cite{AdamsHigdonEtAl12,oden2011grand}.

The need to quantify uncertainties arises in three problem classes within
CSE: (1) The {\em inverse problem}: Given a model, (possibly
noisy) observational data, and any prior knowledge of model parameters
(used in the broadest sense), infer unknown parameters and their
uncertainties by solving a statistical inverse problem. (2) The {\em
prediction (}or {\em forward) problem}: Once model parameters and
uncertainties have been estimated from the data, propagate the
resulting probability distributions through the model to yield
predictions of quantities of interest with quantified
uncertainties. (3) The {\em optimization problem}: Given an objective
function representing quantities of interest
and decision variables (design or control) that can be manipulated to
influence the objective, solve the optimization problem governed by
the stochastic forward problem to produce optimal values of these
variables.

These three classes can all be thought of as
``outer problems,'' since they entail repeated solution of the
deterministic forward problem, namely, the ``inner problem,'' for
different values of the stochastic parameters. However, viewing the
stochastic inverse, forward, and optimization problems merely as
drivers for repeated execution of the deterministic forward problem is
prohibitive, especially when these problems involve large
complex models (such as with PDEs) 
and high-dimensional stochastic parameter spaces (such as when
parameters represent discretized fields). Fundamentally, what ties
these three problems together is the need to explore a parameter space
where each point 
entails a large-scale forward model solve.
Randomly exploring this space with conventional Monte
Carlo methods is intractable.

The key to overcoming the severe
mathematical and computational challenges in bringing UQ to CSE models
is to recognize that beneath the apparently high-dimensional
inversion, prediction, and optimization problems lurk much
lower-dimensional manifolds that capture the maps from
(inversion/design/control) inputs to outputs of interest.
Thus, these problems are characterized by their much smaller intrinsic
dimensions. Black-box methods developed as generic tools are incapable
of exploiting the low-dimensional structure of the operators
underlying these problems. Intensive research is ongoing to develop UQ
methods that exploit this structure, for example by
sparsity-capturing methods, reduced-order models, randomized algorithms, and
high-order derivatives.
This research integrates and cross-fertilizes ideas from statistics, computer
science, numerical analysis, and applied mathematics, while exploiting
the structure of the specific inverse, prediction, and optimal design
and control operators in the context of target CSE problems.
The new high-fidelity, truly predictive science that has emerged within
the past decade and that addresses these issues can be referred to as
{\em predictive CSE}.
Ultimately, the ability to account for uncertainties in CSE
models will be essential in order to bring the full power of modeling and
simulation to bear on the complex decision-making problems facing
society.

%% file: optimization.tex

\medskip
\noindent {\bf Optimization and optimal control.}
\label{sec:optimization}
The past several decades have seen the development of theory and
methods for optimizing systems governed by large-scale CSE
models, typically involving ODEs or PDEs. Such problems usually
take the form of optimal control, optimal design, or inverse
problems. Among other reasons, these problems are challenging because,
upon discretization, the ODEs or PDEs result in very high-dimensional
nonlinear constraints for the optimization problem; exploiting the
structure of these constraints is essential in order to make the solution of the
differential equation-constrained optimization problem tractable. The
optimization problems are made further challenging by additional
inequality constraints, in the form of state or control
constraints.

We identify five areas where research is needed. First, the
successes of the methods mentioned above must be extended to more
complex (multiscale/multiphysics) state equations. Second, methods must be developed
that overcome the curse of dimensionality associated with discrete
decision variables. Third, enhancing scalability of methods developed for nonsmooth
objectives or constraints is critical for a number of large-scale applications.
Fourth, the increasing interest in optimization of
systems governed by stochastic ODEs or PDEs necessitates the
creation of a new class of stochastic optimization methods that can
handle problems with very high-dimensional constraints and 
random variables. This requires advancing mathematical models
of optimal control under uncertainty and risk.
Fifth, optimization methods are needed that
can rigorously employ reduced models for much (or even all) of the
optimization as surrogates for the underlying high-fidelity ODE/PDE
models when the latter become particularly expensive to
solve.

%% file: high-order-AMR.tex

\medskip
\noindent {\bf Highly accurate discretizations and adaptive grid refinement.}
\label{sec:high-order-AMR}
Complex simulations on realistic geometries challenge the capabilities of traditional single-scale methods that utilize quasi-equidistant grids and methods of fixed order. Furthermore, hardware developments favor methods with high arithmetic complexity and low memory footprint.
The natural answer to these challenges is to focus on accurate discretizations, often of high or variable order, in combination with full spatial adaptivity; substantial developments have occurred in both of these key technologies.

Discontinuous Galerkin methods are a prominent example of a family 
of discretizations that have come to fruition as flexible and robust modeling tools. 
Challenges remain when strong discontinuities occur in the solution, and uniformly high-order accurate limiters are needed that remain robust for complex problems on complex grids. Another class of successful methods is the high-order accurate essentially non-oscillatory (ENO) schemes and weighted essentially non-oscillatory (WENO) schemes. 
Isogeometric methods, based on rational splines used in geometry descriptions, have positioned themselves as a powerful tool for fluid-structure problems and complex multiphysics problems. 
Another area of important advances is the development of mimetic finite-volume and finite-element methods. These
methods aim to retain physically relevant geometric and conservation 
properties of the PDE operators on the discrete level,
often leading to important advantages in terms of stability, accuracy, and convergence.
Numerical methods for fractional and stochastic differential equations 
are further areas of current interest,
along with meshless methods, boundary element methods, and radial basis function methods.

Solution-adaptive grid refinement is a key methodology for tackling problems with widely varying spatial scales. 
Diverse methods have been developed that may use block-based, patch-based, or cell-based approaches, with or without overlap. 
Research in error estimation 
has a long tradition, but the essential question in computational 
practice---how discretization and iteration error affect each 
other---remains an open problem. Other research topics include the dynamic 
coarsening and refinement on ever-increasing processor numbers
as well as local timestepping for time-dependent problems.
In order to ensure the usefulness of solution-adaptive simulations on future
computers, dynamic load-balancing must be developed to work in scenarios with millions
of processors.

The development of highly accurate methods during the past decade has addressed a number of key bottlenecks, and substantial advances have been made that enable the robust solution of large multiscale and multiphysics problems using advanced computing platforms. One central challenge that touches on all existing techniques is the development of robust and efficient linear and nonlinear solvers and preconditioning techniques in these more complex scenarios. 
The development significantly trails that of low-order solvers,
and progress is needed to fully benefit from high-order
methods in large-scale problems of scientific and industrial relevance.

%% file: model-order-reduction.tex

\medskip
\noindent {\bf Approximation: simplified, surrogate, and reduced models.}
A significant body of CSE research concerns the conception, analysis, scalable implementation, and application of approximation methods. These methods introduce systematic approximations to the computational model of the system at hand---thus reducing the computational cost of solving the model, while at the same time effecting rigorous control of the resulting error.
While computational efficiency is important in all applications, it is a critical consideration in two particular settings.
First is the \emph{real-time} setting, which translates into (often severe) constraints on analysis time, and sometimes also constraints on memory and bandwidth. Real-time applications span many fields, such as process control, aircraft onboard decision-making, and visualization. Second is the \emph{many-query} setting, in which an analysis (i.e., a forward simulation) must be conducted many times. Examples of many-query applications include optimization, uncertainty quantification, parameter studies, and inverse problems.

Approximation methods can take many forms. The power of multilevel approximations, such as hierarchies of spatial discretizations in a multigrid solver, has long been recognized as an elegant means for exploiting the mathematical structure of the problem and thus obtaining computational speedups. More recently, multilevel approximations have been shown to have similar benefits in uncertainty quantification, such as through the multilevel Monte Carlo method.  Another class of approximation methods seeks to approximate the high-fidelity model of interest by deriving a surrogate model. This surrogate model may take many forms---by applying simplifying physical assumptions, through data-fit regression and interpolation approaches, or via projection of the high-fidelity model onto a low-dimensional subspace. Projection-based model reduction has become a widely used tool, particularly for generating efficient low-cost approximations of systems resulting from parameterized PDEs. Data-driven surrogate models, often drawing on the tools of machine learning, are also starting to see wider development within the CSE community. 
In the context of sampling and integration, Quasi-Monte Carlo methods are explored as alternatives to Monte Carlo methods for high-dimensional problems.
With the drive toward CSE applications of increasing complexity, ensuring the computational tractability of CSE methods and algorithms is becoming increasingly important but also increasingly more challenging. In this regard, an important area of future research involves extending rigorous approximation methods to problems with high dimensionality and with challenging nonlinear and multiphysics behavior. 
\label{sec:model-order-reduction}

%% file: randomized-algorithms.tex

\medskip
\noindent {\bf Randomized algorithms.}
\label{sec:randomized-algorithms}

The field of design and analysis of scalable randomized algorithms is
experiencing rapid growth. 
In the context of science and engineering problems, randomized
algorithms find applications in the numerical solution of PDEs,
model reduction, optimization, inverse
problems, UQ, machine learning, and network
science.

Many classical sequential and parallel algorithms are based on
randomization, for example in discrete mathematics (sorting, hashing,
searching, and graph analysis problems), computational geometry
(convex hulls, triangulation, nearest-neighbors, clustering), and
optimization and statistics (derivative-free solvers, sampling, random
walks, and Monte Carlo methods). 
In the past two decades, however, significant
developments have broadened the role of randomized algorithms by
providing new theoretical insights and significant opportunities for
research in CSE. A first example is compressive sensing, in which
under certain conditions one can circumvent the Nyquist density
sampling barrier
by exploiting sparsity.  A second example is the factorization of
matrices and tensors by using probing,
in which one exposes and exploits global or block-hierarchical low-rank structures
that may exist in the target matrices and tensors.
Such algorithms can be used for accelerating algebraic
operations with tensors and matrices such as multiplication and
factorization. A third example is randomized gradient descent
(especially for nonsmooth problems) for large-scale optimization,
which is popular in both large-scale inverse problems and machine
learning applications.
A fourth example is sampling for computational geometry and signal analysis, such as sparse FFTs and
approximate nearest-neighbors.

Deep connections exist among these apparently different
problems. Their analysis requires tools from functional and matrix
analysis, high-dimensional geometry, numerical linear algebra,
information theory, and probability theory. These algorithms achieve
dimension reduction by exploiting sparsity and low-rank
structures of the underlying mathematical objects and exposing these
structures using ingenious sampling. 
Despite the success of these new
algorithms, however, several challenges remain: extensions to nonlinear
operators and tensors, algorithmic and parallel scalability for large-scale problems,
and development of software libraries for high-performance computing systems.

%% file: multibody.tex

\medskip
\noindent {\bf Multibody problems and mesoscopic methods.} 
\label{sec:multibody}
Simulating a large number of interacting objects 
is among the most important and computationally challenging problems in CSE.
Applications range from simulations of very large objects such as stars and
galaxies to very small objects such as atoms and molecules.
In between is the human scale, dealing with moving objects such as pedestrians,
or the mesoscopic scale, dealing with granular objects such as blood cells
that are transported through microfluidic devices.
Multibody and particle-based modeling, for example, in the form of the discrete element method,
is rapidly gaining relevance since the models can capture behavior that cannot be represented by 
traditional PDE-based methods and since
its high computational cost can now be accommodated by
parallel supercomputers. When short-range interactions dominate,
efficient parallel implementations can be constructed by suitable data structures 
and a dynamic partitioning of the computational domain. 
Long-range interactions (such as gravity or electrostatic forces)
are even more challenging since they inevitably require a global data exchange 
that can be realized efficiently only by advanced parallel hierarchical algorithms such as 
multipole methods, FFTs, or multigrid methods.

Mesoscopic methods are based on a kinetic modeling paradigm,
with the lattice Boltzmann method and smoothed particle hydrodynamics
being particularly successful representatives.
Mesoscopic algorithms are derived in a 
nonclassical combination of model development and discretization
using the principles of statistical physics. 
This results in an algorithmic structure that
involves explicit timestepping and only nearest-neighbor communication,
typically on simple grid structures.
Kinetic methods are particularly successful when
employed in a multiphysics context. 
Here the underlying particle paradigm and statistical physics nature
enable new and powerful approaches to model interactions, as, for example, when using
the so-called momentum exchange method for fluid-structure interaction.
Challenges lie in improving the algorithms and their parallel implementation,
in particular by improving stability, analyzing and controlling errors,
reducing timestep restrictions, deriving better boundary conditions,
developing novel multiphase models, and incorporating mesh adaptivity.

%% file: multiphysics.tex
\medskip
\noindent {\bf Multiscale and multiphysics models.} 
\label{sec:multiphysics}
Because of increasing demands that simulations capture all relevant
influences on a system of interest, 
multiscale and multiphysics simulations are becoming essential for
predictive science.

A multiscale model of a physical system finds use when important
features and processes occur at multiple and widely varying physical
scales within a problem.  
In many multiscale models, a solution to the
behavior of the system as a whole is aided by computing the solution
to a series of subproblems within a hierarchy of scales. At each level
in the hierarchy, a subproblem focuses on a range of the
physical domain appropriate to the scale at which it operates.
Important advances have been made in this area, but challenges
remain, such as atomistic-continuum coupling and problems without a separation
of scales.
The coupling of scales can also lead to powerful algorithms;
we note that some of the most successful methods for simulating
large systems, such as the multigrid method,
owe their efficiency to using the interaction of multiple scales.

A multiphysics system consists of multiple coupled components, each
component governed by its own principle(s) for evolution or
equilibrium.
Coupling individual simulations
may introduce stability, accuracy, or robustness limitations that are
more severe than the limitations imposed by the individual
components. Coupling may occur in the bulk, over interfaces, or over a narrow buffer zone.
In typical approaches one attempts to uncouple dynamics
by asymptotics and multiscale analysis that eliminates stiffness
from mechanisms that are dynamically irrelevant to the goals of the simulation.
Numerical coupling strategies range from loosely coupled Gauss-Seidel
and operator-splitting approaches to tightly coupled Newton-based
techniques~\cite{KeyesMcInnesWoodwardEtAl13}. 
Recent progress includes aspects of problem formulation, 
multiphysics operator decomposition, discretization,
meshing, multidomain interfaces, interpolation, partitioned 
timestepping, and operator-specific preconditioning.

Next-generation advances require
further mathematical analysis and software design to ensure that splitting and coupling
schemes are accurate, stable, robust, and consistent and are implemented correctly. 
Programming paradigms and mathematics both need to be
revisited, with attention to less-synchronous algorithms employing
work stealing, so that different physics components can complement
each other in cycle and resource scavenging without interference.

%% file: hpc-cse.tex
\subsection{CSE and High-Performance Computing -- Ubiquitous Parallelism}
\label{sec:hpc-cse}
\pagebudget{3}
\team{Uli:Voedivin, Keyes, Jimack}

The development of CSE and high-performance computing are closely interlinked.
The rapid growth of available compute power drives CSE research toward ever more complex
simulations in ever more disciplines.
In turn, new paradigms in HPC present
challenges and opportunities that drive future CSE research and education.

\subsubsection{Symbiotic Relationship between CSE and HPC}
HPC and CSE are intertwined in a symbiotic relationship:
HPC technology enables breakthroughs in CSE
research, and leading-edge CSE applications are the main drivers
for the evolution of supercomputer 
systems~\cite{scales03,EESI-generic,ECP-website,wissenschaftsrat-hpc,wissenschaftsrat-simulation,Rosner10,ASCACTopTen2014}.
Grand-challenge applications exercise computational technology at its limits and beyond.
The emergence of CSE as a fundamental pillar
of science has become possible because
computer technology can deliver
sufficient compute power to create effective computational models.
Combined with the tremendous algorithmic advances
(see Figure~\ref{Fig:Algorithmic-Moore}),
these computational models can deliver predictive power and serve
as a basis for important decisions.

On the other hand, the computational needs of CSE applications
are a main driver for HPC research.
CSE applications often require closely interlinked systems,
where not only is the aggregate instruction throughput essential,
but also the tight interconnection between components:
CSE often requires high-bandwidth and low-latency interconnects.
These requirements differentiate scientific computing in CSE
from other uses of information-processing systems.
In particular, many CSE applications cannot be served efficiently
by weakly coupled networks as in grid computing or
generic cloud computing services.

\subsubsection{Ubiquitous Parallelism: A Phase Change for CSE Research and Education}

Parallelism is fundamental for extreme-scale computing, but the
significance of parallelism goes much beyond the topics arising in
supercomputing. All modern computer architectures are parallel, even
those of moderate-sized systems and desktop machines. Since
single-processor clock speeds have stagnated, any further increase in
computational power can be achieved only by a further increase in
parallelism, with ever more complex hierarchical and heterogeneous
system designs. High-performance computing architectures already are
incorporating large numbers of parallel threads, possibly reaching a
billion by the year 2020.

Future mainstream computers for science and engineering will not be
accelerated versions of current architectures but, instead, smaller
versions of extreme-scale machines. In particular, they will inherit
the node and core architecture from the larger machines. Although
computer science research is making progress in developing techniques
to make architectural features transparent to the application
developer, doing so remains an ongoing research effort. Moreover, the
technological challenges in miniaturization, clock rate, bandwidth
limitations, and power consumption will require deep and disruptive
innovations, dramatically increasing the complexity of software
development.

Parallel computing in its full breadth thus has become a central and
critical issue for CSE. Programming methodologies must be adapted, not
only for the extreme scale but also for smaller parallel
systems. Efficient and sustainable realizations of numerical libraries
and frameworks must be designed. Furthermore, for high-end
applications, the specifics of an architecture must be explicitly
exploited in innovative algorithm designs. To meet these demands
requires a dramatic phase change for both CSE research and education.

\subsubsection{Emergent Topics in HPC-Related Research and Education}
\begin{figure}[h]
\centering
\includegraphics[width=0.7\textwidth]{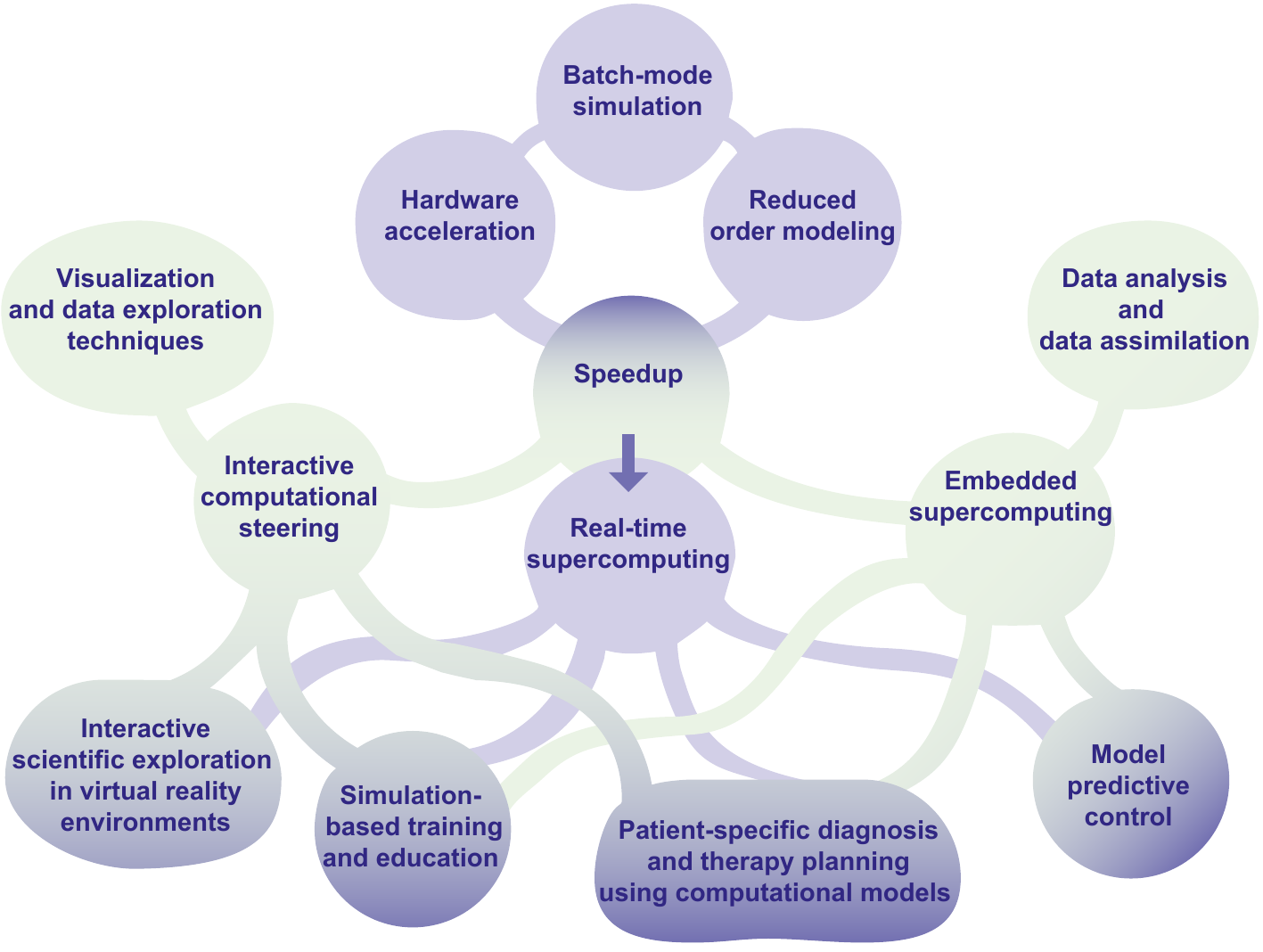}
\caption{Some emerging developments based on real-time or embedded HPC methodology for CSE applications.}
\label{fig:emergent-cse}
\end{figure}

{\bf Extending the scope of CSE through HPC technology.}
Low-cost 
computational power, becoming available through
accelerator hardware such as graphic processing units (GPUs),
increasingly is enabling nontraditional uses of HPC technology for CSE.
One significant opportunity arises in real-time and embedded supercomputing.
Figure~\ref{fig:emergent-cse} illustrates a selection of possible
future development paths, many of which involve
advanced interactive computational steering
and/or real-time simulation.
Once simulation software can be used in real time,
it can also be used for training and education.
A classical application is the flight simulator, but the
methodology can be extended to many other situations where humans operate complex technical objects
and where systematic training on a simulator may save time and money as well as increase preparedness for emergency situations.
Further uses of fast, embedded CSE systems include the development of
simulators for the modeling of predictive control systems and for patient-specific biomedical diagnosis.
The development of these
emerging CSE applications, shown
in Figure~\ref{fig:emergent-cse}, will require a focused investment 
in parallel computing research and education.

As another example, extreme-scale computing will enable mesoscale simulation
to model the collection of cells that make up a human organ or a large collection of particles 
{\em directly}, without resorting to averaging approaches.
The simulation of granular material has tremendous practical
importance.
The examples range from the transport and processing of
bulk materials and powders in industry
to the simulation of avalanches and landslides.
The potential that arises with the advent of powerful supercomputers
can be seen when realizing that exascale means $10^{18}$ but that
a human has ''only'' around $10^{11}$
neurons and $10^{13}$ red blood cells, and that
the 3D printing of a medical implant may require to process
$10^8$ individual grains of titanium alloy.
Thus, extreme-scale computation may open the route to modeling techniques where each cell or grain is represented individually.  This gives rise to
research directions that are out of reach on conventional computer systems
but that will exceed the predictive power of continuum models
for such simulation scenarios.
In order to exploit these opportunities, new simulation methods must be devised, 
new algorithms invented, and new modeling paradigms formulated.
New techniques for validation and verification are needed.
Fascinating opportunities in fundamental research arise that go 
far beyond just developing new material laws and 
increasing the mesh resolution in continuum models.
\medskip

\noindent
{\bf  Quantitative performance analysis of algorithms and software.}
The advent of exascale and other performance-critical applications
requires that CSE research and education address the performance
abyss between the traditional mathematical assessment of
computational cost and the
implementation of algorithms on current computer systems.
The traditional cost metrics based on counting
floating-point operations fail increasingly to correlate with the truly relevant
cost factors, such as time to solution or energy consumption.
Research is necessary in order to quantify
more complex algorithmic characteristics, such as memory footprint and
memory access structure (cache reuse, uniformity of access,
utilization of block transfers, etc.), processor utilization,
communication, and synchronization requirements. These effects must be
built into better cost and complexity models.

Furthermore, the traditional approach to
theory in numerical analysis
provides only an insufficient basis to quantify the efficiency of
algorithms and software, since many theorems are only qualitative and 
leave the constants unspecified.
Such mathematical results, although themselves rigorous,
permit only heuristic---and thus often misleading---predictions of real computational
performance.
Thus much of current numerical analysis research must be
fundamentally extended to become better guiding principles for
the design of efficient simulation methods in practice. 
\medskip

\noindent
{\bf Performance engineering and co-design.} 
In current CSE practice, performance models are used to analyze existing 
applications for current and future computer systems; 
but the potential of performance analysis techniques is rarely used 
as a systematic tool for designing, developing, and implementing
CSE applications.
In many cases an a priori analysis 
can be used to determine the computational resources that
are required for executing a specific algorithm.
Where available, such requirements (e.g., flops, memory, memory bandwidth, network bandwidth)
should be treated as nonfunctional goals
for the software design.

Required here is a fundamental shift from the current practice
of treating performance as an a posteriori diagnostic assessment to recognizing performance as
an a priori design goal.
This step is essential because when performance criteria are considered
too late in the design process,
fundamental decisions (about data structures and algorithms) cannot be revised,
and improvements are then often limited to an unsatisfactory
code-tuning and tweaking.
The idea of an a priori treatment of performance goals in scientific software engineering
is related to the {\em co-design} paradigm
and has become a new trend for developing next-generation 
algorithms and application software systems.
The nations of the G-8 have instituted regular meetings to strategize
about the formidable task of porting to emerging exascale
architectures the vast quantity of software on which computational
science and engineering depend.  These co-design efforts were codified
in the 2011 International Exascale Software Project Roadmap~\cite{dongarra_ijhpca2011}.

\medskip

\noindent
{\bf Ultrascalability.} 
For the foreseeable future all growth in compute power
will be delivered through increased parallelism.
Thus, we must expect that high-end applications will reach
degrees of parallelism of up to $10^9$ within a decade.
This situation poses a formidable challenge to the design
and implementation of algorithms and software.
Traditional paradigms of bulk-synchronous operation are likely to face significant performance obstacles.
New communication-avoiding algorithms must be designed and analyzed.
Many algorithms permit increased asynchronous executions enabling processing to continue even if a small number of processors stay behind; but this is a wide-open area of research because it requires a new look at data dependencies,
exploiting task-based parallelism, and possibly also nondeterministic execution schedules.
Additionally, system software must be extended to permit the efficient
and robust implementation of such asynchronous algorithms.
\medskip

\noindent
{\bf Power wall.}
The increasing aggregate compute capacity in highly parallel computers both on the desktop and on the supercomputer promises to enable the simulation of
problems with unprecedented size and resolution.
However, electric power consumption per data element is
not expected to drop at the same rate. This creates a {\it power wall},
and power consumption is emerging as one of the fundamental
bottlenecks of large-scale computing.
How dramatic this will be for computational science can be seen from the following simple example.

Assume that the movement of a single word of data can be estimated by 
$1$ NJoule ($=10^{-9}$ Joule) of energy~\cite{SachsYelickEtAl2011}.
If we now assume that a specific computation deals with 
$N=10^{9}$ entities (such as mesh nodes or particles),
then using an $O(N^2)$ algorithm to transfer $N \times N$ data items for an all-to-all interaction,
such as computing a pairwise distance,
will cause an energy dissipation of $10^{18}$ NJoule $\approx 277$ kWh.
Assuming a petaflop computer (which may become available on the desktop in the coming decade),
we could in theory execute 
the $N^2$ operations in 1000~seconds. 
However, a cost of 277 kWh for the naively implemented data movement
will require more than 1 MW of sustained power intake.
Clearly such power levels are neither feasible nor affordable in a standard environment.

The situation gets even more dramatic 
when we transition to {\em tera}scale problems with $N=10^{12}$ on supercomputers.
Then the same all-to-all data exchange will dissipate an enormous 277 GWh,
which is equivalent to the energy
output of a medium-size nuclear fusion explosion.
Clearly, in application scenarios of such scale, a global data movement with
$O(N^2)$ complexity must be
classified as practically impossible.
Only with suitable hierarchical algorithms
that dramatically reduce the complexity 
can we hope to
tackle computational problems of such size.
This kind of bottleneck must be addressed in both research and education.
In the context of CSE, the power wall becomes
primarily a question of designing the most efficient algorithm in terms of
operations {\em and also} the data movement.
Additionally, more energy-efficient hardware systems need to be
developed.

\medskip

\noindent
{\bf Fault tolerance and resilience.}
With increasing numbers of functional units and cores and with continued miniaturization,
the potential for hardware failures rises.
Fault tolerance on the level of a system can be reached only by redundancy,
which drives the energy and investment cost. At this time
many algorithms used in CSE are believed to have good potential for so-called algorithm-based fault tolerance. That is, the
algorithm either is naturally tolerant against certain faults
(e.g., still converges to the correct answer, but perhaps more slowly)
or can be augmented to compensate for different types of failure (by exploiting specific features of the data structures and the algorithms, for example).
Whenever there is hierarchy, different levels and pre-solutions
can be used for error detection and circumvention.
At present, many open research questions arise from these considerations,
especially when the systems, algorithms, and applications are studied in combination.

%% file: data-cse.tex
\subsection{CSE and the Data Revolution: The Synergy between Computational Science and Data Science}
\label{sec:data-cse}
\pagebudget{3}
The world is experiencing an explosion of digital data.
Indeed, since 2003, new data has been growing at an annual rate that exceeds the data contained in all previously created documents.
The coming of extreme-scale computing and data acquisition from high-bandwidth experiments across the sciences is creating a phase change.
The rapid development of networks of sensors and
the increasing reach of the Internet and other digital networks in our connected society
create new data-centric analysis applications in broad areas of science,
commerce, and technology \cite{baker2010data,oden2011grand,jahanian2013testimony}.
The massive
amounts of data offer tremendous potential for generating new
quantitative insight, not only in the natural sciences and
engineering, where they enable new approaches such as data-driven
scientific discovery and data-enabled uncertainty quantification, but also
in almost all other areas of human activity.
For example, biology and
medicine have increasingly become quantitative sciences over the past
two or three decades, aided by the generation of large data sets. Data-driven
approaches are also starting to change the social sciences,
which are becoming more quantitative \cite{king2014restructuring}.

\subsubsection{CSE and the Data Revolution: The Paradigms of Scientific Discovery}
CSE has its roots in the third paradigm of scientific discovery -- computational modeling, and it drives scientific and technological progress in conjunction with the first two paradigms, theory and experiment, making use of first-principles models that reflect the laws of nature. These models may, for example, include the PDEs of fluid mechanics and quantum physics
or the laws governing particles in molecular dynamics.
The advent of big data is sometimes seen as 
enabling a fourth paradigm of scientific discovery \cite{hey2009fourth}, 
in which the sheer amount of data combined with statistical models
leads to new 
analysis methods in areas where first-principles models do not exist (yet) or are inadequate.

Massive amounts of data are indeed creating a sea change in scientific discovery.
In third-paradigm CSE applications (that are based on first-principles models) big data leads to
tremendous advances: it enables revolutionary methods of data-driven discovery,
uncertainty quantification, data assimilation, optimal design and control, and, ultimately,
truly predictive CSE. At the same time, in fourth-paradigm approaches big data
makes the scientific method of quantitative, evidence-based analysis applicable to
entirely new areas where until recently quantitative data and models were mostly
nonexisting. 
The fourth paradigm also enables new approaches in the physical sciences
and engineering, for example, for pattern finding in large amounts of observational data.
Clearly, CSE methods and techniques have an essential role to play
in all these quantitative endeavors enabled by big data.

\subsubsection{Role of Big Data in CSE Applications}
In core application areas of CSE~\cite{oden2011grand}, our ability to produce data is rapidly outstripping our ability to use it.
With exascale data sets, we will be creating far more data than we can explore in a lifetime with current tools.
Yet, exploring these data sets is the essence of new paradigms of scientific discovery.
Thus, one of the greatest challenges is
to create new theories, techniques, and software that can be used to
understand and make use of this rapidly growing data
for new discoveries and advances in science and engineering.
For example, the CSE focus area of
uncertainty quantification aims at characterizing and managing the uncertainties
inherent in the use of CSE models and data. To this end, new methods are
being developed that build on statistical techniques such as Monte Carlo methods, Bayesian inference, 
and Markov decision
processes. While these underlying techniques have broad applications
in many areas of data science, CSE efforts tend to have a special
focus on developing efficient structure-exploiting computational techniques at scale, with
potential for broad applicability in other areas of data analytics
and data science. 
Data assimilation methods have over several decades evolved into crucial techniques
that ingest large amounts of measured data into large-scale computational models for
diverse geophysical applications such as weather prediction and hydrological forecasting. 
Large amounts of data are also a crucial component in other
CSE focus areas, such as validation and verification, reduced-order
modeling, and analysis of graphs and networks. Also, enormous potential
lies in the emerging model-based interpretation of patient-specific data from
medical imaging for diagnosis and therapy planning.
CSE techniques to
address the challenges of working with massive data sets include
large-scale optimization, linear and nonlinear solvers, inverse problems, stochastic methods, scalable
techniques for scientific visualization, and high-performance parallel
implementation.

Exploiting large amounts of data is having a profound influence in many areas of CSE
applications. The following paragraphs describe some striking examples.

Many {\bf geoscience systems} are characterized by complex behavior
coupling multiple physical, chemical, and/or biological processes over
a wide range of length and time scales. Examples include earthquake
rupture dynamics, climate change, multiphase reactive subsurface
flows, long-term crustal deformation, severe weather, and mantle
convection. The uncertainties prevalent in the mathematical and
computational models characterizing these systems have made
high-reliability predictive modeling a challenge. However, the
geosciences are at the cusp of a transformation from a largely
descriptive to a largely predictive science. This is driven by
continuing trends: the rapid expansion of our ability to instrument
and observe the Earth system at high resolution, sustained
improvements in computational models and algorithms for complex
geoscience systems, and the tremendous growth in computing power.

The problem of how to estimate unknown parameters (e.g., initial
conditions, boundary conditions, coefficients, sources) in complex
geoscience models from large volumes of observational data is
fundamentally an inverse problem. Great strides have been made in the
past two decades in our ability to solve very large-scale geoscience
inverse problems, and many efforts are under way to parlay these
successes for deterministic inverse problems into algorithms for
solution of Bayesian inverse problems, in which one combines possibly
uncertain data and models to infer model parameters and their
associated uncertainty. When the parameter space is large and the
models are expensive to solve (as is the usual case in geoscience
inverse problems), the Bayesian solution is prohibitive.

However, advances in large-scale UQ algorithms in recent years
\cite{AdamsHigdonEtAl12} are beginning to make feasible the use
of Bayesian inversion and Markov chain Monte Carlo methods
to infer parameters and their uncertainty in
large-scale complex geoscience systems from large-scale satellite
observational data. Two examples are global ocean modeling
and continental ice sheet modeling.
Continued advances in UQ algorithms,
Earth observational systems, computational modeling, and HPC systems
over the coming decades will lead to more sophisticated geoscience
models capable of much greater fidelity. These
in turn will lead to a better understanding of Earth dynamics as well
as improved tools for simulation-based decision making for critical
Earth systems.

Big data methods are revolutionizing
the related fields of {\bf chemistry} and {\bf materials science}, in a
transformation that is illustrative of those sweeping all of science,
leading to successful transition of basic science into practical tools
for applied research and early engineering design.
Chemistry and materials science are both
mature computational disciplines that through advances in theory,
algorithms, and computer technology are now capable of increasingly
accurate predictions of the physical, chemical, and electronic
properties of materials and systems. The equations of quantum mechanics
(including Schr\"odinger's, Dirac's, and density functional representations)
describe the electronic structure of solids and molecules that controls many
properties of interest, and statistical mechanics 
 must be employed to
incorporate the effects of finite temperature and entropy. These are
forward methods---given a chemical composition and approximate
structure, one can determine a nearby stable structure and
compute its properties. To design new materials or chemical
systems, however, one must solve the inverse problem---what is the system
that has specific or optimal properties?  Moreover, the system must be
readily synthesized, inexpensive, and thermally and chemically stable
under expected operating conditions. Breakthrough progress has recently been
made in developing effective constrained search and optimization
algorithms for precisely this purpose
\cite{ceder2013stuff},
with this process
recognized in large funding initiatives such as the multiagency U.S. Materials
Genome Initiative \cite{materialsGenome}.  This success has
radically changed the nature of computation in the field.  Less than
ten years ago most computations were generated and analyzed by a human,
whereas now 99.9\% of computations are
machine generated and processed as part of automated searches that are
generating vast databases with results of millions of
calculations to correlate structure and function
\cite{materialsProject,molecularSpace}.
In addition to opening important new challenges in robust and
reliable computation, the tools and workflows of big data are now
crucial to further progress.

\medskip
\begin{tcolorbox}[title=CSE Success Story: Visual Analytics Brings Insight to Terabytes of Simulation Data]
\begin{wrapfigure}{R}{0.7\textwidth}
\vspace{-0.15in}
\includegraphics[width=0.7\textwidth]{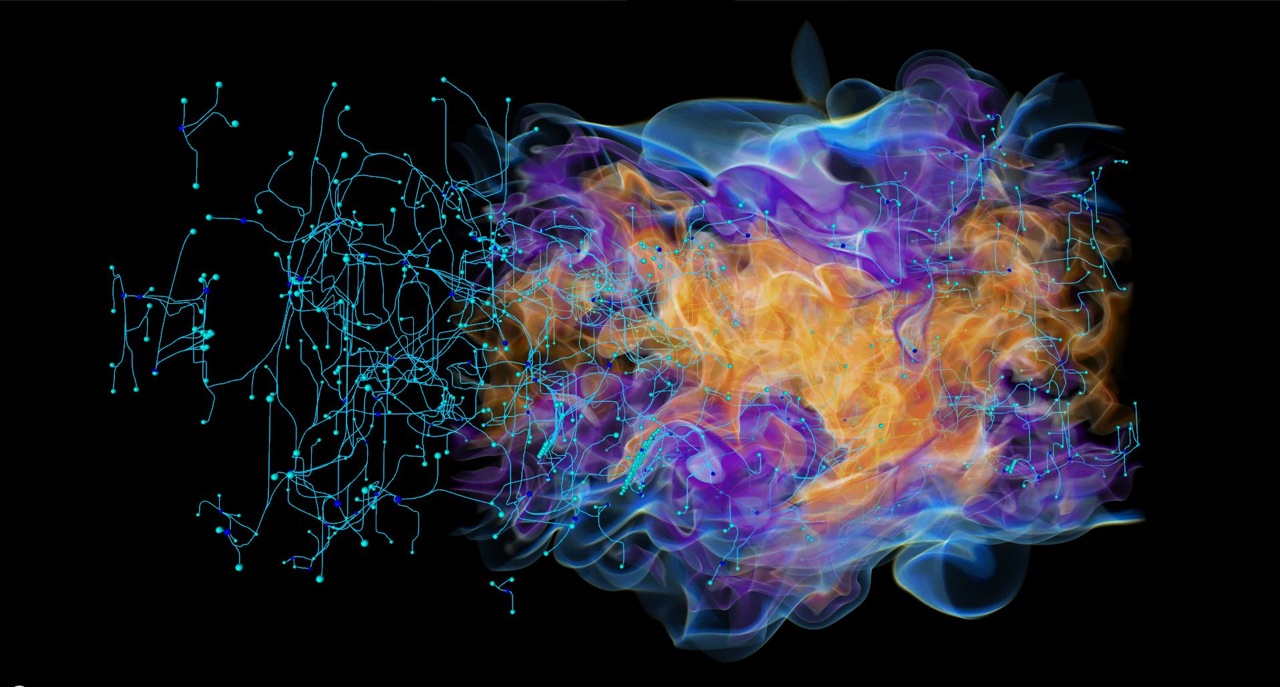}
\end{wrapfigure}
New techniques are being developed that allow scientists to sift through terabytes of simulation data in order to gain important new insights from science and engineering simulations on the world's largest supercomputers. The figure shows a visualization of a topological analysis and volume rendering of one timestep in a large-scale, multi-terabyte combustion simulation. The topological analysis identifies important physical features (ignition and extinction events) within the simulation, while the volume rendering allows viewing the features within the spatial context of the combustion simulation.\protect\footnotemark
\end{tcolorbox}

\footnotetext{Simulation by J. Chen, Sandia National Laboratories; visualization by the Scientific Computing and Imaging Institute, University of Utah.}

In {\bf scientific visualization}, new techniques are being developed to give visual insight
into the deluge of data that is transforming scientific research. 
Data analysis and visualization are key technologies for enabling advances in simulation and data-intensive science, as well as in several domains beyond the sciences.  Specific big data visual analysis challenges and opportunities include in situ interactive analysis, user-driven data reduction, scalable and multilevel hierarchical algorithms, representation of evidence and uncertainty, heterogeneous data fusion, data summarization and triage for interactive queries, and analysis of temporally evolved features \cite{johnson2006nih,johnson2007visualization,wong2012top}.

Computation and big data also meet in {\bf characterization of physical
material samples} using techniques such as X-ray diffraction and adsorption,
neutron scattering, pytchography, transmission electron, and atomic
microscopes. Only for essentially perfect crystals or simple
systems can one directly invert the experimental data and determine
the structure from measurements.
Most real systems, typically with nanoscale features and no long-range
order, are highly underdetermined \cite{billinge2010nano}.  Reliable
structure determination requires fusion of multiple experimental data
sources (now reaching multiple terabytes in size) and computational
approaches.  Computation provides a forward simulation
(e.g., given a structure, determine what spectrum or diffraction pattern results),
and techniques from uncertainty quantification are among those proving
successful in making progress.

\subsubsection{Synergy between Computational Science and Data Science}
Big data is transforming the fabric of society, in areas that go
beyond research in the physical sciences and engineering
\cite{jahanian2013testimony}. Data analytics aims at extracting information
from large amounts of data in areas as diverse as business intelligence, cybersecurity, social
network recommendation, and government policy. Analysis of the data is
often based on statistical and machine learning methods from data
science. Similar to CSE, data science is built on fundamentals from
mathematics and statistics, computer science, and domain knowledge;
and hence it possesses an important synergy with CSE.

The paradigm of scalable algorithms and implementations
that is central to CSE and HPC is also relevant to
emerging trends in data analytics and data science.
Data analytics is quickly moving in the direction of mathematically more sophisticated analysis algorithms and parallel implementations. CSE will play an
important role in developing the next generation of parallel
high-performance data analytics approaches that employ
descriptions of the data based on physical or phenomenological models
informed by first principles,
with the promise of extracting valuable insight from the data that crucially goes beyond
what can be recovered by statistical modeling alone.
Important mathematical and algorithmic advances in areas such as optimization,
randomized algorithms, and approximation are currently driven by problems in
machine learning and deep learning.

HPC supercomputers and cloud data centers serve different needs and are
optimized for applications that have fundamentally different characteristics.
Nevertheless, they face challenges that have many
commonalities in terms of extreme scalability,
fault tolerance, cost of data movement, and power management.
The advent of big data has spearheaded new large-scale distributed
computing technologies and parallel programming models
such as MapReduce, Hadoop, Spark, and Pregel,
which offer innovative approaches to scalable high-throughput
computing, with a focus on data locality and fault tolerance.
These frameworks are finding applications in CSE problems, for
example, in network science; and large-scale CSE methods,
such as advanced distributed optimization algorithms, are
increasingly being developed and implemented in these
environments.
In many applications, the need for distributed computing arises from the
sheer volume of the data to be processed and analyzed,
and, similar to the discussions on HPC in Section \ref{sec:hpc-cse},
the growing levels of parallelism in computer architectures
require software in distributed machine learning systems such as
TensorFlow to be highly parallel.
Extensive potential exists for cross-fertilization of
ideas and approaches between extreme-scale HPC
and large-scale computing for data analysis.
Economy-of-scale pressures will contribute
to a convergence of technologies for computing
at large scale.

Overall, the analysis of big data
requires efficient and scalable mathematics-based algorithms executed
on high-end computing infrastructure, which are core CSE competencies
that translate directly to big data applications. CSE
education and research must foster the important synergies with
data analytics and data science that are apparent in a variety of
emerging application areas.

%% file: software-cse.tex
\subsection{CSE Software}
\label{sec:software-cse}
\pagebudget{3}

CSE software ecosystems provide fundamental and pervasive
technologies that connect advances in applied mathematics, computer
science, and core disciplines of science and engineering
for advanced modeling, simulation, discovery, and analysis.
We discuss the importance and scope of CSE software, the increasing challenges in CSE software development and sustainability, and the future CSE software research agenda.

\subsubsection{Importance and Scope of CSE Software}
Software is an essential
product of CSE research when complex models of reality are cast into algorithms; moreover,
the development of efficient, robust, and
sustainable software is at the core of CSE.
The CSE agenda for research includes the systematic design and
analysis of (parallel) software, its accuracy, and its computational
complexity (see Section~\ref{sec:hpc-cse}).
Beyond this, CSE research must deal with the
assessment of computational cost on the relevant hardware platforms,
as well as with criteria such as flexibility, usability,
extensibility, and interoperability.
Software that contributes to modeling, simulation, and analysis is only
part of the software required in CSE.  Equally important are operating
systems, programming models, programming languages, compilers,
debuggers, profilers, source-to-source translators, build systems,
dynamic resource managers, messaging systems, I/O systems, workflow
controllers, and other types of system software that support
productive human-machine interaction.
Software in this wider sense
also includes the infrastructure necessary to support a CSE research
ecosystem, such as version control, automatic tests for correctness
and consistency, documentation, handbooks, and tutorials.  All this
software is essential for CSE to continue to migrate up computational
scales, and it requires an interdisciplinary community to produce it
and to ensure that it coheres.

While the volume and complexity of scientific software have grown
substantially in recent decades~\cite{GroppHarrisonEtAl2016},
scientific software traditionally has not received the focused
attention it so desperately needs in order to fulfill this key
role as a cornerstone of long-term CSE
collaboration and scientific
progress~\cite{SoftwareProductivityWorkshop14,HerouxAllenEtAl2016,Hettrick2016}.
Rather, ``software has evolved organically and
inconsistently, with its development and adoption coming largely as
by-products of community responses to other targeted
initiatives'' \cite{KeyesTaylor2011}.

\subsubsection{Challenges of CSE Software}
The community faces increasing challenges in CSE software design, development,
and sustainability as a result of the confluence of disruptive changes in
computing architectures and demands for more complex simulations.  New
architectures require fundamental algorithm and software
refactoring, while at the same time enabling new ranges of
modeling, simulation, and analysis.
\medskip

\smallskip
\noindent
{\bf New science frontiers: increasing software demands.}  CSE's
continual push toward new capabilities that enable predictive science
dramatically affects how codes are designed, developed, and used.
Software that incorporates multiphysics and multiscale modeling, capabilities
beyond interpretive simulations (such as UQ
and design optimization), and coupled data
analytics presents a host of difficulties not faced in
traditional contexts, because of the compounded complexities of code
interactions~\cite{KeyesMcInnesWoodwardEtAl13,HerouxAllenEtAl2016,SoftwareProductivityWorkshop14}.
A key challenge is
enabling the introduction of new models, algorithms, and data
structures over time---that is, balancing competing goals of interface
stability and software reuse with the ability to innovate
algorithmically and develop new approaches.

\smallskip
\noindent
{\bf Programmability of heterogeneous architectures.}
Designing and developing CSE software
to be sustainable are challenging
software engineering tasks,
not only in the extreme scale, but also in conventional applications
that run on standard hardware.
The best software architecture is often
determined by performance considerations, and it is a high
art to identify kernel routines that can serve
as an internal interface to a software performance layer.
While the optimization of the kernel routines inevitably requires detailed knowledge of a
specific target machine, the design the software architecture must 
support optimizations not only for current but also for future generations of computer systems.
Such long term sustainability of software is essential to amortize the high cost of
developing complex CSE applications, but it
requires a deep understanding of computer architecture
and its interplay with algorithms.

All modern computers are hierarchically structured. This structure, in turn, creates the need to
develop software with the hierarchy and the architecture in mind,
often using a hybrid combination of different languages and tools.
For example, a given application may utilize MPI on the system level,
OpenMP on the node level, and special libraries or low-level intrinsics to exploit core-level vectorization.
Newer techniques from computer science, such as automatic program generation, annotations,
and domain-specific languages, may eventually help reduce the gap between
real-life hardware structures and model complexity.

This complexity will need to be managed, and to some extent alleviated, in the future.
For example, the development of new and improved unifying languages, combined with the tools to select appropriate algorithms for target architectures and to implement these algorithms automatically,
may ease the burden on CSE software developers.
Such tools are topics
of current research and are therefore far from reaching the level of
maturity required to support large-scale development.
Consequently, CSE developers must currently rely on an approach that includes hardware-optimized libraries; or they must
master the complexity---typically in larger teams where members can specialize---by undertaking explicitly hardware-aware development.
This task is even more complex when accelerators,
such as GPUs, are to be used.

\smallskip
\noindent
{\bf  Composability, interoperability, extensibility, portability.}
As CSE applications increase in
sophistication, no single person or team possesses the expertise and
resources to address all aspects of a simulation.  Interdisciplinary
collaboration using software developed by independent groups becomes
essential.  CSE researchers face daunting challenges in developing,
deploying, maintaining, extending, and effectively using libraries,
frameworks, tools, and application-specific infrastructure.

Practical difficulties in collaborative research software stem from
the need for composable and interoperable code with support for
managing complexity and change as architectures, programming models,
and applications continue to advance. Challenges include
coordinating the interfaces between components that need to
interoperate and ensuring that multiple components can be
used side by side without conflicts between programming models and
resources. Even more difficult challenges arise with the need to
exchange or control data between components, where many issues center
on ownership and structure of the data on which components act.
Moreover, good software must be extensible, to meet not only requirements known at
the time of its design but also unanticipated needs that change over time.
Software must also be portable across target architectures, including
laptops, workstations, and moderately sized clusters for much of the
CSE community.  Even researchers who employ the full resources of
emerging extreme-scale machines typically develop and test their code
first on laptops and clusters, so that portability across this
entire spectrum is essential.

\medskip
\begin{tcolorbox}[title=CSE Success Story:
Numerical Libraries Provide Computational Engines for Advanced CSE Applications]
\begin{wrapfigure}{R}{0.58\textwidth}
\vspace{-0.15in}
\includegraphics[width=0.58\textwidth]{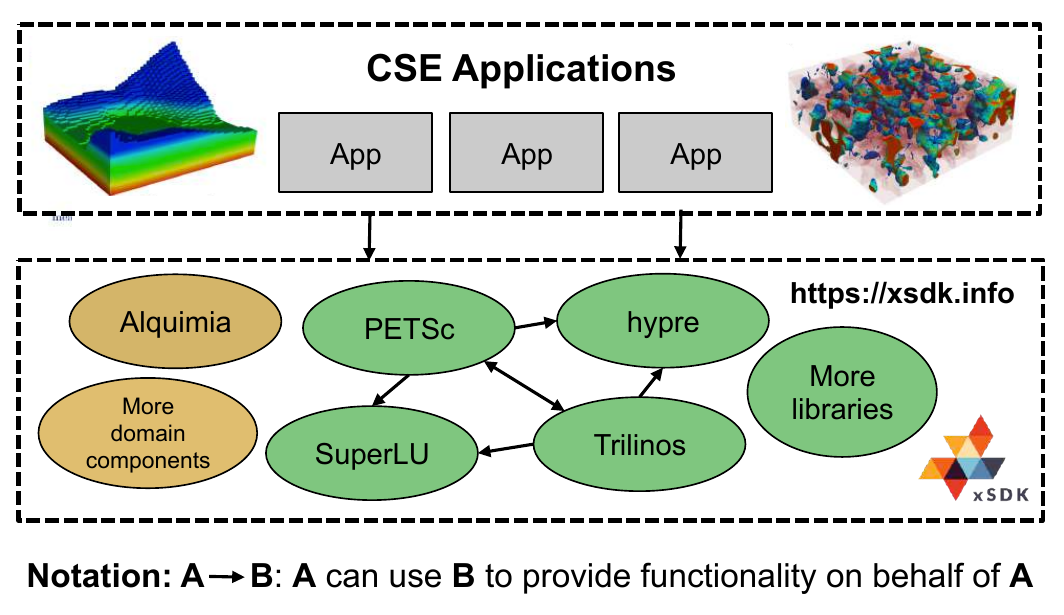}
\end{wrapfigure}

Community collaboration and support are essential in
driving and transforming how large-scale, open-source software
is developed, maintained, and used.  Work is under way in
developing the Extreme-scale Scientific Development Kit
(xSDK),\protect\footnotemark ~which is improving the interoperability,
portability, and sustainability of CSE libraries and application
components. The vision of the xSDK is to provide the foundation of
an extensible scientific software ecosystem
developed by diverse, independent teams throughout the
community, in order to improve the quality, reduce the cost, and
accelerate the development of CSE applications.
The xSDK incorporates, for example, the high-performance numerical
libraries hypre, PETSc, Sundials, SuperLU, and Trilinos, which are
supported by the U.S. Department of Energy and
encapsulate cutting-edge algorithmic advances
to achieve robust, efficient, and scalable performance on
high-performance architectures.
These packages provide the
computational engines for thousands of
advanced CSE applications, such as
hydrology and biogeochemical cycling simulations using PFLOTRAN and
coupled 3D microscopic-macroscopic steel simulations (far left and far right images, respectively, in the ``CSE Applications'' box in the diagram).
For example, this multiphase steel application, which
uses nonlinear and linear FETI-DP domain decomposition methods (in PETSc) and algebraic
multigrid (in hypre),
demonstrates excellent performance on the entire Blue Gene/Q at the
J\"{u}lich Supercomputing Centre (JUQUEEN, 458K cores) and the
Argonne Leadership Computing Facility (Mira, 786K cores).
\end{tcolorbox}

\footnotetext{
Information on xSDK available via \url{https://xsdk.info}.
PFLOTRAN simulations by G. Hammond (Sandia National Laboratories).
Multiphase steel simulations described in A.~Klawonn, M. Lanser, and O. Rheinbach,
SIAM J Sci Comput 37(6), C667-C696, 2015; image courtesy of
J\"org Schr\"oder, Universit\"at Duisburg-Essen.}

\subsubsection{Software as a Research Agenda for CSE}
CSE software ecosystems, which support scientific research in
much the same way that a light source or telescope does, require a
substantial investment of human and capital resources, as well as
basic research on scientific software productivity so that the resulting software
artifacts are fully up to the task of predictive simulations and
decision support.
Scientific software often has a much longer lifetime than does hardware;
in fact, software frequently outlives the teams that originally create it.
Traditionally, however, support for software has generally been
indirect, from funding sources focused on science or engineering
outcomes, and not the software itself. This circumstance---sporadic,
domain-specific funding that considers software only secondary to the
actual science it helps achieve---has caused huge difficulties,
not only for sustainable CSE software artifacts, but also for
sustainable CSE software careers.
This in turn has increasingly led to a mismanagement of
research investment, since scientific software as an
important CSE research outcome is rarely leveraged to its full potential.

Recent community reports express the imperative to firmly embrace
the fundamental role of open-source CSE software as a valuable research product
and cornerstone of CSE collaboration and thus to increase direct investment in
the software itself, not just as a byproduct of other
research~\cite{pitac05,GroppHarrisonEtAl2016,HerouxAllenEtAl2016,Hettrick2016,SoftwareProductivityWorkshop14,KeyesTaylor2011}.
The past decade has seen the development of many successful community-based open-source CSE software projects
and community science codes.
Aided by advances in supporting technology such as version control, bug tracking, and online collaboration, these projects leverage broad communities to develop free software with features at the leading edge of algorithmic research.
Examples in the area of finite-element methods include the deal.II, Dune, and FEniCS projects; and similar efforts have made tremendous contributions in many other areas of CSE.

\smallskip
\noindent
{\bf Reproducibility and sustainability.}
CSE software often captures the essence of research results.
It must therefore be considered equivalent to other scientific
outcomes and must be subjected to equivalent quality assurance procedures.
This requirement in turn means that criteria such as the reproducibility
of results~\cite{StoddenEtAl2013} must be given higher priority and that CSE software must be
more rigorously subjected to critical evaluation
by the scientific community.
Whenever a research team is faced with increased expectations for
independent review of computational results, the team's interest in
improved software methodologies increases commensurately.  In fact, it
is not too strong to say that the affordability and feasibility of
reproducible scientific research are directly proportional to the
quality and sustainability of the software.
Community efforts are beginning to address issues in
software sustainability, or the ability to maintain the scientifically useful
capability of a software product over its intended life span,
including understanding and modifying a software product's behavior to reflect
new and changing needs and technology~\cite{wssspe:home,HerouxAllenEtAl2016,Hettrick2016}.
Work is needed to determine value metrics for CSE software that fully
acknowledge its key role in scientific progress; to increase rewards
for developers of open-access, reliable, extensible, and sustainable
software; and to expand career paths for expert CSE software
developers.

\smallskip
\noindent
{\bf Software engineering and productivity.}
The role of software ecosystems as foundations for CSE discoveries
brings to the forefront issues of
software engineering and productivity,
which help address reproducibility and sustainability.
Software productivity expresses the effort, time, and cost of
developing, deploying, and maintaining a product having needed software
capabilities in a targeted scientific computing
environment~\cite{HerouxAllenEtAl2016,SoftwareProductivityWorkshop14}.
Work on software productivity focuses on improving the quality, decreasing
the cost, and accelerating the delivery of scientific applications,
as a key aspect of improving overall scientific productivity.
Software engineering, which can be defined as ``the application of
a systematic, disciplined, quantifiable approach to the development,
operation, and maintenance of software'' \cite{IEEE-glossary}, is central to any effort
to increase CSE software productivity.

While the
scientific community has much to learn from the mainstream software
engineering community,
CSE needs and environments are in combination sufficiently unique so as to
require fundamental research specifically for scientific software.
In particular, scientific software domains require extensive academic
background in order to understand how software can be designed,
written, and used for CSE investigations. Also, scientific
software is used for discovery and insight,
and hence requirements (and therefore all other phases of the
software lifecycle) are frequently changing.

Consequently, CSE software ecosystems and processes urgently require
focused research and substantial investment.
Another pressing need is education on
software engineering and productivity methodologies that are
specifically tailored to address the unique aspects of CSE, in the
contexts of both academic training and ongoing professional
development (see Section \ref{sec:education}).
With respect to many of these issues, CSE software research is still nascent,
since these themes have
been largely neglected in the evolution of the field to date.
As stated in a recent NITRD report~\cite{HerouxAllenEtAl2016},
``The time is upon us to address the growing challenge of
software productivity, quality, and sustainability that imperils the whole endeavor of
computation-enabled science and engineering.''

%% file: predictive-cse.tex

\subsection{Emergence of Predictive Science}
\label{sec:predictive-cse}

The advances in CSE modeling, algorithms, simulation, big data analytics, HPC, and
scientific software summarized in this document
all have the overarching goal of achieving truly predictive science capabilities.
Scientific experimentation and theory, the classical paradigms of the scientific method,
both strive to describe the physical world.
However, high-fidelity predictive capabilities can
be achieved only by employing numerical computation.
Predictive science now lies at the core of the new CSE discipline.

CSE draws its predictive power from mathematics, statistics, and the natural sciences
as they underlie model selection, model calibration, model validation,
and model and code verification, all in the presence of uncertainties.
Ultimately CSE must also include the propagation of uncertainties through the forward problem
and the inverse problem, to quantify the uncertainties of the outputs that are the target goals of the simulation.
When actual computer predictions are used for critical decisions,
all of these sources of uncertainty must be taken into account.

Current algorithms and methods for coping with these issues have
their roots in the mathematics and statistics of the past century and earlier.
In order to deal with the complexities of predictive modeling, however, new models, algorithms, and methodologies are needed. Their development, analysis, and implementation constitute the new research agenda for CSE.
Achieving these goals will require substantial research efforts and significant breakthroughs.

What predictive science and therefore what CSE will eventually be
is not yet fully understood.
We may see the coastline of
the ``continent of predictive science'' ahead of us, but we still have to explore
the whole mass of land that lies behind this coastline.
We can already clearly see, however,
that CSE and the transition to predictive science
will have a profound impact on education,
on how scientific software is developed,
on research methodologies,
and on the design of tomorrow's computers.

%% file: education.tex
\section{CSE Education and Workforce Development}
\label{sec:education}

With the many current and expanding opportunities for the CSE field, there is a growing demand for CSE graduates and a need to expand CSE educational offerings. This need includes CSE programs at both the undergraduate and graduate levels, as well as continuing education and professional development programs. In addition, the increased presence of digital educational technologies provides an exciting opportunity to rethink CSE pedagogy and modes of educational delivery.

\subsection{Growing Demand for CSE Graduates}

Industry, national laboratories, government, and broad areas of academic research are making more use than ever before of simulations, high-end computing, and simulation-based decision-making.  This trend is apparent broadly across domains---for example, energy, manufacturing, finance, and transportation are all areas in which CSE is playing an increasingly significant role, with many more examples across science, engineering, business, and government. Research and innovation, both in academia and in the private sector, are increasingly driven by large-scale computational approaches.
A National Council on Competitiveness report points out that high-end computing plays a ``vital role in driving private-sector competitiveness ... all businesses that adopt HPC consider it indispensable for their ability to compete and survive" \cite{ConC2008}.
With this significant and increased use comes a demand for a workforce versed in technologies necessary for effective and efficient mathematics-based computational modeling and simulation. There is high demand for graduates with the interdisciplinary expertise needed to develop and/or utilize computational techniques and methods in order to advance the understanding of physical phenomena in a particular scientific, engineering, or business field and to support better decision-making \cite{GlotzerKimEtAl2009}.

As stated in a recent report on workforce development by the U.S. Department of Energy Advanced
Scientific Computing Advisory Committee  \cite{ASCAC2014}, ``All large
DOE national laboratories face workforce recruitment and retention
challenges in the fields within Computing Sciences that are relevant
to their mission. ...
There is a growing national demand for graduates in Advanced Scientific Computing Research-related Computing Sciences
that far exceeds the supply from academic institutions. Future
projections indicate an increasing workforce gap.''
This finding was based on a number of reports,
including one from the High End Computing Interagency Working Group \cite{HEC-IWG2013}
stating: ``High end computing (HEC) plays an important role
in the development and advanced capabilities of many of the products,
services, and technologies that are part of our everyday life. The
impact of HEC on the agencies of the federal government, on the quality
of academic research, and on industrial competitiveness is substantial
and well documented.  However, adoption of HEC is not uniform, and to
fully realize its potential benefits we must address one of the most
often cited barriers: lack of HEC skills in the workforce."
Additional workforce and education issues are discussed in \cite{HerouxAllenEtAl2016}.

The U.S. Department of Energy has for 25 years been investing in the
Computational Science Graduate Fellowship~\cite{CSGF:website} program
to prepare approximately 20 Ph.D.\ candidates per year for
interdisciplinary roles in its laboratories and beyond.  Fellows take
at least two graduate courses in each of computer science, applied
mathematics, and an application from science or engineering requiring
large-scale computation, in addition to completing their degree
requirements for a particular department.  They also spend at least
one summer at a DOE laboratory in a CSE internship and attend an
annual meeting to network with their peers across other institutions.
This program has been effective in creating a sense of community for
CSE students that is often lacking on any individual traditionally
organized academic campus.

In order to take advantage of the transformation that high-performance
and data-centric computing offers to industry, the critical factor is a
workforce versed in CSE and capable of developing the algorithms,
exploiting the compute platforms, and designing the analytics that
turn data with its associated information into knowledge to act.
This is the case for large companies that have traditionally had in-house simulation capabilities and may have dedicated CSE-focused groups to support a wide range of products; it is also increasingly the case for small- and medium-sized companies with more specialized products and a critical need for CSE to support their advances in research and development.
In either case, exploiting emerging computational tools requires the critical thinking and
the interdisciplinary background that is prevalent in CSE training~\cite{oden2011grand}.
The CSE practitioner has the expertise to apply computational tools
in uncharted areas, often applying previous domain-specific understanding.
The CSE practitioner also has the analytical skills to tease out the problems that often are
encountered when commercial enterprises seek to design new products,
develop new services, and create novel approaches from the wealth of data
available.
While often a member of a team of others from varying
disciplines, the CSE practitioner is the catalyst driving the change that industry seeks in order
not only to remain competitive but also to be first to market, providing the
necessary advantage to thrive in a rapidly evolving technological
ecosystem.

\subsection{Future Landscape of CSE Educational Programs}

CSE educational programs are needed in order to create
professionals who meet this growing demand and who support the
growing CSE research field. These include CSE programs at both the
undergraduate and graduate levels, as well as continuing education and
professional development programs. They also include programs that are
``CSE focused'' and those that follow more of a ``CSE infusion''
model. The former include programs that have CSE as their primary
focus (e.g., B.S., M.S., or Ph.D. in computational science and engineering),
while the latter include
programs that embed CSE training within another degree structure
(e.g., a minor, emphasis, or concentration in CSE complementing a major
in mathematics, science, or engineering or a degree in a specific
computational discipline such as computational finance or
computational geosciences). In fact, interdisciplinary quantitative and
computer modeling skills are quickly becoming indispensable for any
university graduate, not only in the physical and life sciences, but also in the social sciences.
Universities must equip their graduates with these skills.
Information about a variety of CSE educational programs can be found online~\cite{SIAG-CSE:wiki,CSE-GraduateProgramSurvey2012}.

{\bf Undergraduate education.}  At the undergraduate level, the breadth and depth of topics covered in CSE degrees
depends on the specific degree focus. However, the following high-level topics are
important content for an undergraduate program:
\begin{enumerate}
\item Foundations in mathematics and statistics, including calculus, linear algebra, mathematical analysis, ordinary and partial differential equations, applied probability, stochastic processes, and discrete mathematics
\item Simulation and modeling, including conceptual, data-based, and physics-based models, use of simulation tools, and assessment of computational models
\item Computational methods and numerical analysis, including errors, solutions of systems of linear and nonlinear equations, Fourier analysis, interpolation, regression, curve fitting, optimization, numerical differentiation and integration, Monte Carlo methods, statistical inference, numerical methods for ODEs, and numerical methods for PDEs
\item Computing skills, including compiled high-level languages, algorithms (numerical and nonnumerical), elementary data structures, analysis of algorithms and their implementation, parallel programming, scientific visualization, awareness of computational complexity and cost, and use of good software engineering practices including version control
\end{enumerate}
Feedback from the community has noted an increasing demand for CSE graduates trained at the bachelor's level, with particular note of the increased opportunities at small- and medium-sized companies. A report from the SIAM Working Group on CSE Undergraduate Education further develops foundations for directions in undergraduate CSE education \cite{education2011undergraduate}.

{\bf Graduate education.}  At the graduate level, again the breadth and depth of topics covered depends on the specific degree focus. In the next section, we make specific recommendations in terms of a set of learning outcomes desired for a CSE graduate program.
We also note the growing importance of and demand for terminal master's degrees, which can play a large role in fulfilling the industry and national laboratory demand for graduates with advanced CSE skills.
All CSE graduates should possess the attributes of having a solid foundation in mathematics; an understanding of probability and statistics; a grasp of modern computing, computer science, programming languages, principles of software engineering, and high-performance computing; and an understanding of foundations of modern science and engineering, including biology. These foundations should be complemented by deep knowledge in a specific area of science, engineering, mathematics and statistics, or computer science. CSE graduates should also possess skills in teamwork, multidisciplinary collaboration, and leadership.
A valuable community project would be to collect resources to assist early-career researchers in advancing skills to support CSE collaboration and leadership.

{\bf Continuing and professional education.} A third area of educational programs is that of continuing and professional education. Opportunities exist for SIAM or other institutions
to engage with industry to create and offer short courses, including those that target general CSE skills for the non-CSE specialist as well as those that target more advanced skills in timely opportunity areas (such as parallel and extreme-scale computing,
CSE-oriented software engineering, and computing with massive data).
Often one assumes that much of the workforce for industry in CSE will
come at the postgraduate level; increasingly, however, industry needs
people who have an understanding of CSE even at the undergraduate
level in order to realize the full potential growth in a rapidly
expanding technological workplace.
Future managers and leaders in
business and industry must be able to appreciate the skill
requirements for the CSE professional
and the benefits that accrue from CSE. Continuing
education can play a role in fulfilling this need. The demand for
training in CSE-related topics exists broadly among graduate
students and researchers in academic institutions and national
laboratories, as evidenced by the growing number of summer schools
worldwide, as well as short courses aimed at the research community.
For example, the Argonne Training Program on Extreme-Scale
Computing~\cite{atpesc:website} covers key topics that CSE researchers
must master in order to develop and use leading-edge applications on
extreme-scale computers.  The program targets early-career researchers
to fill a gap in the training that most computational scientists
receive and provides a more comprehensive program than do typical short
courses.\footnote{Videos and slides of lectures are available via the ATPESC
website \url{http://extremecomputingtraining.anl.gov}.}
Continuing education also has an important role to play in addressing the challenge of changing computer architectures---skills developed around optimizing algorithms for today's machines might become obsolete within the lifetime of a student's professional career.
Lastly, we note that the recent creation of the SIAM Activity Group on Applied Mathematics Education
 represents another opportunity for collaboration to pursue some of these ideas in continuing and professional education.

{\bf Institutional structure.}
Because of CSE's intrinsically interdisciplinary nature and its research agenda reaching beyond the traditional disciplines, the development of CSE is often impeded by traditional institutional boundaries.
CSE research and education have found great success over the past
decade in those settings where CSE became a clearly articulated focus of
entire university departments,\footnote{For example, the School of
  Computational Science \& Engineering at the Georgia Institute of
  Technology and the Department of Scientific Computing at Florida State
  University.
  } faculties,\footnote{For example, the
  Division of Computer, Electrical, and Mathematical Sciences and
  Engineering at the King Abdullah University of Science and
  Technology (KAUST).
  }
or large
interdisciplinary centers.\footnote{For example, the Institute for
  Computational Engineering and Sciences at the University of Texas at
  Austin, the Scientific Computing and Imaging Institute at the
  University of Utah, the Cluster of Excellence in Simulation
  Technology at the University of Stuttgart,
  and CERFACS (Centre Européen de Recherche et de
  Formation Avancée en Calcul Scientifique) in Toulouse.}
In many of the newer universities in the world, institutional structures
often develop naturally in line with the CSE paradigm.\footnote{KAUST
  is, again, a good example.}
In other cases, institutional traditions and realities make it more
natural for successful CSE programs to develop within existing
departments\footnote{For example, the master's program in CSE at the
Technische Universit{\"a}t M\"{u}nchen.} or in
cross-departmental\footnote{For example, CSE graduate programs in
  engineering faculties at the University of Illinois at
  Urbana-Champaign, at the Massachusetts Institute of Technology,
  and at the Technische Universit{\"a}t Darmstadt}
  or cross-faculty\footnote{For example, the
  School of Computational Science and Engineering at McMaster
  University, the Institute for Computational and Mathematical Engineering at Stanford University,
  and the master's program in CSE at the Ecole
  Polytechnique Federale de Lausanne.}  initiatives.
  Information about a variety of CSE educational programs can be found online~\cite{SIAG-CSE:wiki,CSE-GraduateProgramSurvey2012}.
In any case, universities and research institutes
will need to implement new
multidisciplinary structures that enable
more effective CSE research and education.
One ingredient appears crucial for success, regardless of the particular institutional structure:
It is critical to have both top-down support from the university administration and ground-up enthusiasm from the faculty.
To fully realize its potential, the CSE endeavor requires its own
academic structures, funding programs, and educational programs.

\medskip
\begin{tcolorbox}[title=CSE Success Story: Computational Modeling and Data Analytics Undergraduate Degree at Virginia Tech]
\begin{wrapfigure}{R}{0.5\textwidth}
\includegraphics[width=0.5\textwidth]{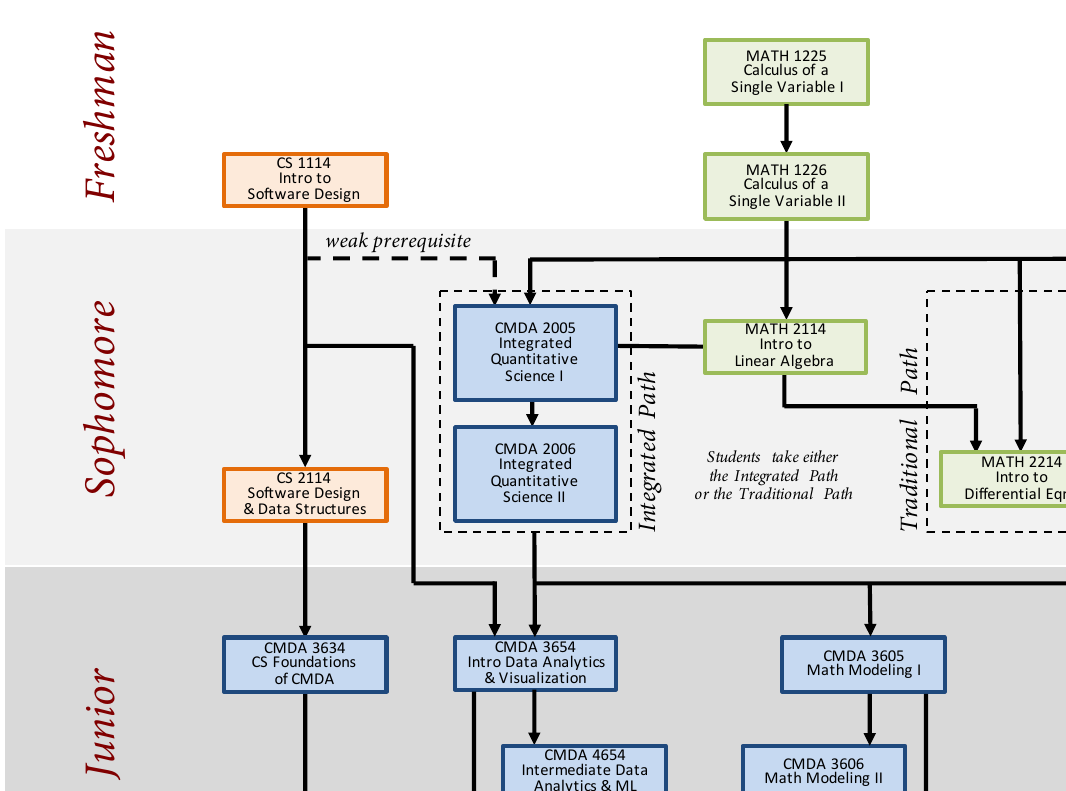}
\end{wrapfigure}
In Spring 2015, Virginia Tech launched a new undergraduate major in Computational Modeling and Data Analytics (CMDA).  The curriculum is a collaboration among the Departments of Computer Science, Mathematics, and Statistics, across the Colleges of Science and Engineering.  The program includes ten new courses specially designed for the major, building skills in dynamical systems and mathematical modeling, statistics and data analytics, and high-performance computing; specialized options in physics and economics are available.  The curriculum intentionally builds an interdisciplinary perspective.  For example, sophomore level multivariable calculus, differential equations, probability, and statistics are taught together in two 6-credit courses (Integrated Quantitative Science), team taught by a mathematician and statistician.  A capstone project course emphasizes leadership, teamwork, communication, and project management skills, as small teams tackle semester-long modeling and analytics challenges from clients.  The new degree program has proven to be popular: total enrollment in Spring 2017 was already above 300 students, with the first class of 22 graduating in May 2017.
\end{tcolorbox}

\subsection{Graduate Program Learning Outcomes}

\input{outcomes-education}

A learning outcome is defined as what a student is expected to be able
to do as a result of a learning activity. In this section, we describe
a set of learning outcomes desired of a student graduating from a CSE
Ph.D. program. We focus on outcomes because they describe the set of
desirable competencies without attempting to prescribe any specific
degree structure. These outcomes can be used as a guide to define a
Ph.D. program that meets the needs of the modern CSE graduate; they can
also play an important role in defining and distinguishing the CSE
identity and in helping employers understand the skills and potential
of CSE graduates.

In Table~\ref{tab:LO}, we focus on the ``CSE Core Researchers and Developers" category in
Figure \ref{Fig:CSE-community}. We distinguish between a CSE Ph.D. with a broadly applicable
CSE focus and a CSE Ph.D. with a domain-driven focus. An example of the
former is a ``Ph.D. in computational science and engineering" while
an example of the latter is a ``Ph.D. in computational
geosciences." The listed outcomes relate primarily to those
CSE-specific competencies that will be acquired through classes. Of
course, the full competencies of the Ph.D. graduate must also include
the more general Ph.D.-level skills, such as engaging deeply in a
research question, demonstrating awareness of research context and
related work, and producing novel research contributions, many of
which will be acquired through the doctoral dissertation.
We note that it would be desirable for graduates of a CSE master's degree program to also achieve most (if not all) of the outcomes in Table~\ref{tab:LO}.
In particular, in educational systems with no substantial classwork component for the Ph.D., the learning outcomes of
Table~\ref{tab:LO} would also apply to the master's or honors degree that may precede the Ph.D.

In the next two subsections, we elaborate on the interaction between CSE education and two areas that have seen considerable change since the design of many existing CSE programs: extreme-scale computing and computing with massive data.

\subsection{Education in Parallel Computing and Extreme-Scale Computing}
\team{Uli: Voedivin, Keyes, Jimack}

Engineers and scientists need to be
better prepared for the age of ubiquitous parallelism (as addressed in Section~\ref{sec:hpc-cse}; see also \cite{HEC-IWG2013,EESI-generic,ECP-website,ASCAC2014,wissenschaftsrat-simulation}).
Parallelism has become the basis for all
computing technology and necessitates a shift
in teaching even the basic concepts.
Simulation algorithms and their
properties have been in the core of CSE education, but now we must
emphasize parallel algorithms.
The focus used to be on abstract notions of accuracy of methods
and the complexity of algorithms;
today it must be shifted to the complexity of parallel
algorithms and the real-life cost of solving a computational
problem---a completely different notion.
Additionally,
the asymptotic complexity and thus algorithmic scalability become more
important when the machines grow larger.
At the same time, the traditional complexity metrics increasingly fail to give guidance
about which methods, algorithms, and implementations are truly efficient.
As elaborated in Sections~\ref{sec:hpc-cse} and \ref{sec:software-cse}, designing
simulation software has become a complex, multifaceted art.
The education of future computational scientists must
address these topics that arise from
the disruptive technology that is dramatically
changing the landscape of computing.

Extreme-scale computing also presents new challenges to education.
Education in programming techniques needs
to be augmented with parallel programming elements
and a distinctive awareness of performance and computational cost.
Additionally the current trends are characterized by a growing complexity in the design
of computer architectures, which are becoming hierarchical and heterogeneous.
These architectures are reflected by
complex and evolving programming models
that should be addressed in a modern CSE education.

Today's extreme scale is tomorrow's desktop. An analogous statement
holds for the size of the data that must be processed and that is
generated through simulations. In education we need to distinguish
between those whose research aims to simulate computationally
demanding problems (see Section~\ref{sec:hpc-cse} and Figure~\ref{fig:emergent-cse})
and the wider class of people who are less driven
by performance considerations. For example, many
computational engineering problems exist in which the models are
not extremely demanding computationally
or in which model reduction
techniques are used to create relatively cheap models.

In defining the educational needs in parallel and high-performance computing
for CSE, we must distinguish
between different intensities. Any broad education in CSE will benefit
from an understanding of parallel computing, simply because sequential
computers have ceased to exist. All students must be trained to
understand concepts such as concurrency, algorithmic complexity, and
its relation to scalability, elementary performance metrics, and systematic
benchmarking methodologies.
In more demanding applications, parallel computing expertise and
performance awareness are necessary and must go significantly beyond the
content of most current curricula. This requirement is equally true in those
applications that may be only of moderate scale but that nevertheless
have high-performance requirements, such as those in real-time
applications or those that require interactivity; see Figure~\ref{fig:emergent-cse}.
Here, CSE education must include
a fundamental understanding of computer architectures and the programming
models that are necessary to exploit these architectures.

Besides classification according to scientific content
and HPC intensity,
educational structures in CSE must also address the wide spectrum of
the CSE community that was described and analyzed in Section~\ref{sec:cse-community}
(see also Figure \ref{Fig:CSE-community}).

\smallskip
\noindent
{\bf CSE Domain Scientists and Engineers -- Method Users.}
Users of CSE technology typically employ dedicated supercomputer
systems and
specific software on these computers; they
usually do not program HPC systems from scratch.
Nevertheless, they need to understand the systems and the software
they use, in order to achieve leading-edge scientific results.
They must be capable of extending the existing applications, if needed,
possibly in collaboration with CSE and HPC specialists.

An appropriate educational program for CSE users in HPC
can be organized in courses and tutorials on specific topics such as
are regularly offered by computing centers and other institutions.
These courses are often taught in compact format (ranging from a few hours to a week)
and are aimed at enabling participants to use specific methods and software or specific systems and tools. They naturally are of limited depth, but a wide spectrum of such courses
is essential in order to widen the scope of CSE and HPC technology and to bring it to bear fruit as
widely as possible.

\smallskip
\noindent
{\bf CSE Domain Scientists and Engineers -- Method Developers.}
Developers of CSE technology are often domain scientists or engineers who have specialized in using computational techniques in their original field. They often have decades of experience in computing
and using HPC,
and thus historically they are mostly self-taught.
Regarding the next generation of scientists, students
of the classical fields
(such as physics, chemistry, or engineering) will increasingly want to put stronger
emphasis on computing within their fields.

The more fundamental knowledge that will be needed to competently use the next generation of HPC systems thus cannot be adequately addressed by compact courses as described above. A better integration of these topics into the university curriculum is necessary,
by teaching the use of computational methods as part of existing courses or by offering dedicated HPC- and simulation-oriented courses (as electives) in the curriculum.
An emphasis on CSE and HPC within a classical discipline may be taught
in the form of a selection of courses  that are offered as electives
by CSE or HPC specialists, or---potentially especially attractive---by co-teaching of a CSE specialist jointly with a domain scientist.

\smallskip
\noindent
{\bf CSE Core Researchers and Developers.}
Scientists who work at the core of CSE are classified in two groups according to Figure~\ref{Fig:CSE-community}. {\em Domain-driven} CSE students as well as those focusing on {\em broadly applicable methods} should be expected to spend a significant amount of time learning about HPC and parallel computing topics. These elements must be well integrated into the CSE curriculum. Core courses from computer science (such as parallel programming, software engineering, and computer architecture) may present the knowledge that is needed also in CSE, and they can  be integrated into a CSE curriculum. Often, however, dedicated courses that are
especially designed for students in CSE will be significantly
more effective, since such courses can be adapted to the special prerequisites of the student group
and can better focus on the issues that are
relevant for CSE. Often again co-teaching such courses, labs, or projects may be fruitful, especially when such courses cover several stages of the CSE 
cycle (see Figure~\ref{Fig:CSE-pipeline}).

These three levels of CSE education are naturally interdependent,
but we emphasize that all three levels are relevant and important.
In particular, the problem
of educating the future generation of scientists
in the competent
use of computational techniques cannot be addressed solely by
offering one-day courses on how to use the latest machine in the computing center.

\medskip
\begin{tcolorbox}[title=CSE Success Story: Computer-Aided Engineering in the Automotive Industry]
\begin{wrapfigure}{R}{0.7\textwidth}
\includegraphics[width=0.7\textwidth]{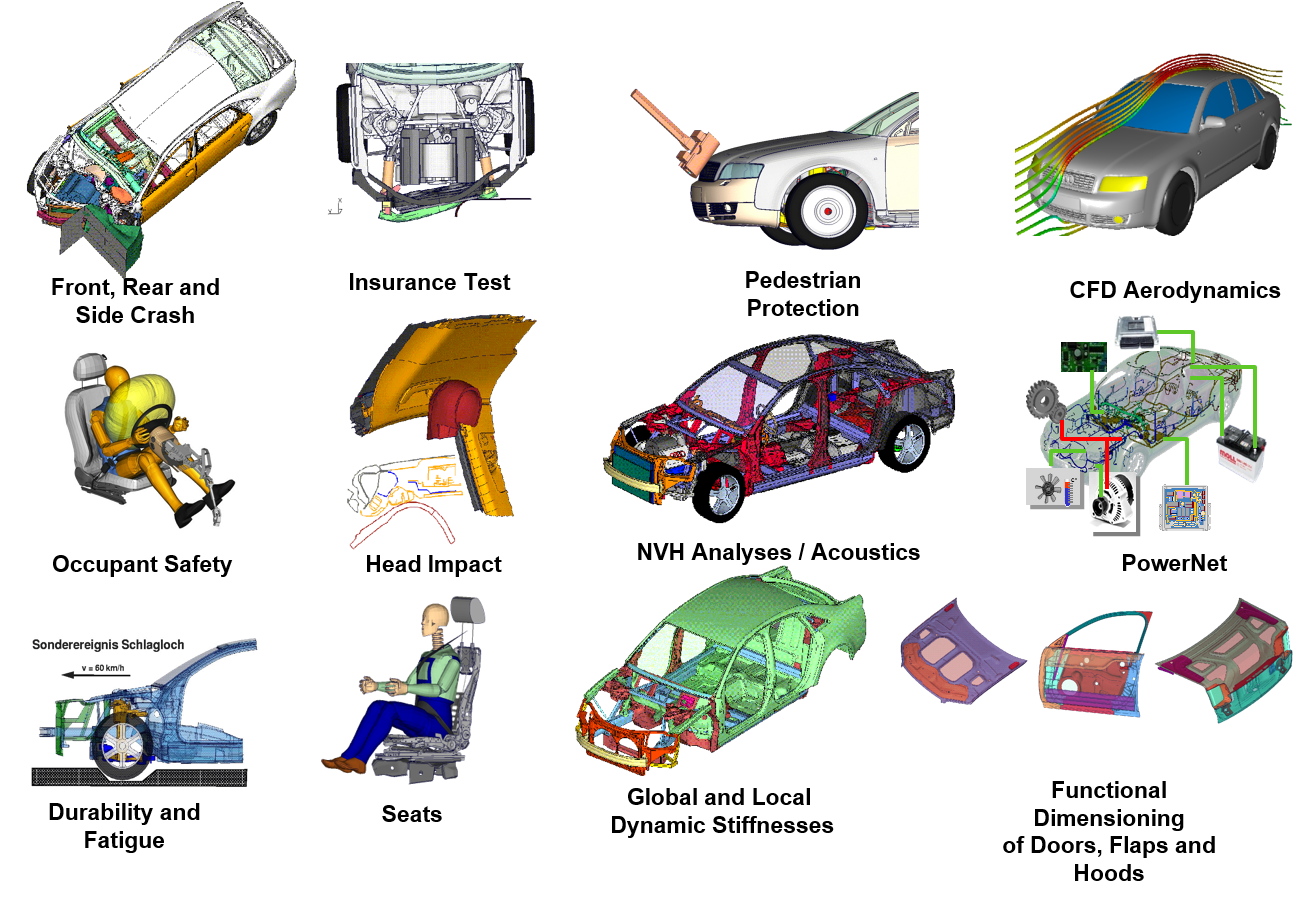}
\end{wrapfigure}
CSE-based simulation using
computer-aided engineering (CAE) methods
and tools has become an indispensable component of developing
advanced products in industry.  Based on mathematical
models (e.g., differential equations and variational principles), CAE
methods such as multibody simulation, finite elements,
and computational fluid dynamics are essential for
assessing the functional behavior of products
early in the design cycle
when physical prototypes are not yet available.
The many advantages of virtual testing
compared with physical testing include flexibility, speed, and cost.
This figure\protect\footnotemark ~shows selected
application areas of CAE in the automotive industry.
CSE provides widely applicable methods and tools.
For example, drop tests of
mobile phones are investigated by applying simulation methods that are
also used in automotive crash analysis.
\end{tcolorbox}

\footnotetext{Figure courtesy of AUDI AG.}

\subsection{CSE Education in Uncertainty Quantification and Data Science}
The rising importance of massive data sets in application areas of
science, engineering, and beyond has broadened the skillset that
CSE graduates may require. For example, data-driven uncertainty
quantification requires statistical approaches that may include tools
such as Markov chain Monte Carlo methods and Bayesian
inference. Analysis of large networks requires skills in discrete
mathematics, graph theory, and combinatorial scientific
computing. Similarly, many data-intensive problems require approaches
from inverse problems, large-scale optimization, machine learning, and
data stream and randomized algorithms.

The broad synergies between computational science and data science
offer opportunities for educational programs. Many CSE competencies
translate directly to the analysis of massive data sets at scale using
high-end computing infrastructure. Computational science and data
science are both rooted in solid foundations of mathematics and
statistics, computer science, and domain knowledge; this common
core may be exploited in educational programs that prepare the
computational and data scientists of the future.

We are already beginning to see the emergence of such programs. For example, the new undergraduate major in ``Computational Modeling and Data Analytics" at Virginia Tech\footnote{\url{http://www.science.vt.edu/ais/cmda/}} includes deep integration among applied mathematics, statistics, computing, and science and engineering applications. This new degree program is intentionally designed \emph{not} to be just a compilation of existing classes from each of the foundational areas; rather, it comprises mostly new classes with new perspectives emerging from the intersection of fields and is team taught by faculty across departments.
Another example is the Data Engineering and Science Initiative at Georgia
Tech.\footnote{\url{http://bigdata.gatech.edu/}, \url{http://www.analytics.gatech.edu/}} Degree programs offered include a one-year M.S. in analytics, and M.S. and Ph.D. programs with a data focus on CSE and biotech fields. These programs are jointly offered by academic units drawn from the colleges of computing, engineering, and business. About a quarter of the courses are offered by the School of CSE, with the focus on computational algorithms and high-performance analytics.
A similar picture is emerging around the world, with interdisciplinary programs that combine data analytics and mathematics-based high-end computing.\footnote{See, e.g., several of the programs listed
at \url{http://www.kdnuggets.com/education/index.html}}

\medskip
\begin{tcolorbox}[title=CSE Success Story: Simulation-Based Optimization of 3D Printing]
\begin{wrapfigure}{R}{0.66\textwidth}
\vspace{-0.15in}
\includegraphics[width=0.33\textwidth]{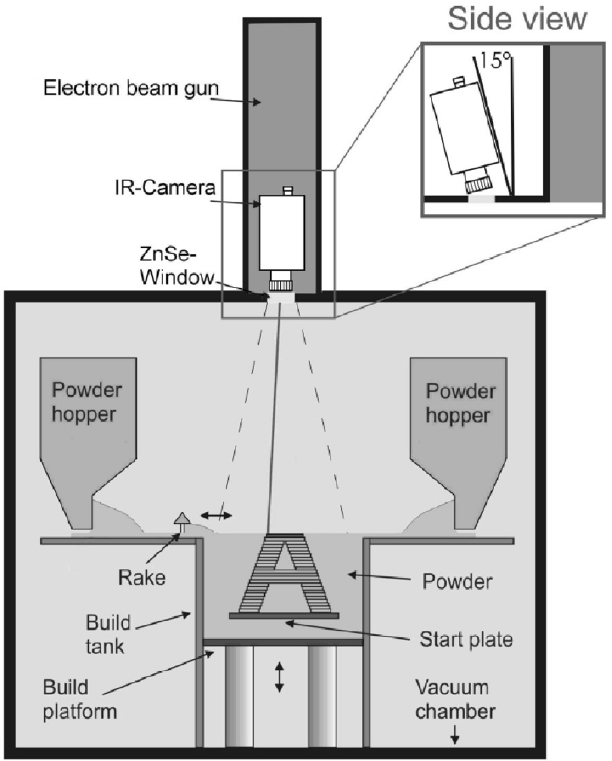}
\hfill
\begin{minipage}[b]{0.33\textwidth}
\includegraphics[width=.9\textwidth]{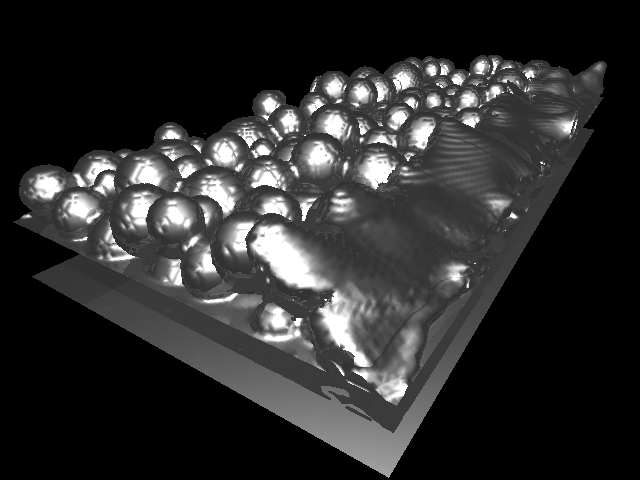}
\\[1ex]
\includegraphics[width=.9\textwidth]{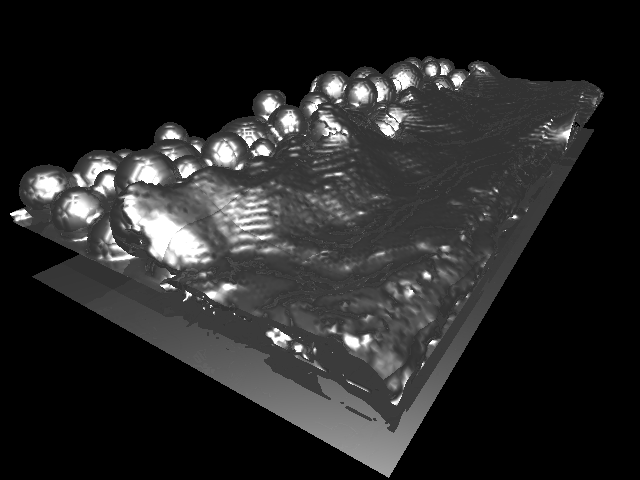}
\end{minipage}
\end{wrapfigure}
CSE researchers have developed advanced models of 3D printing processes, where
thin layers of metal powder are molten by a
high-energy electron beam that welds the
powder selectively to create
complex 3D metal structures
with an almost arbitrary geometry by repeating the process layer by layer.
The two snapshots on the right visualize the effects of
a simulated electron beam that scans over a powder bed in a sequence of
parallel lines. Simulation can be used for designing the electron beam gun, developing
the control system, and generating the powder layer, thereby
accelerating the printing process in commercial manufacturing, for
example, of patient-specific medical implants.
The greatest simulation challenge is to develop numerical models for the
complex 3D multiphysics welding process.
A realistic simulation with physical resolution of a few microns
requires millions of mesh cells and several hundreds of thousands of
timesteps---computational complexity that can be tackled only with
parallel supercomputers and sophisticated software.\protect\footnotemark
\end{tcolorbox}

\footnotetext{
Simulation results from M. Markl, R. Ammer, U. R\"{u}de, and C. K\"{o}rner,
International Journal of Advanced Manufacturing Technology, 78, 239-247, 2015.}

\subsection{Software Sustainability, Data Management, and Reproducibility}
As discussed in Section~\ref{sec:software-cse},
simulation software is becoming increasingly complex and often
involves many developers, who may be geographically distributed
and who may enter or leave the project at different times.
Education is needed on issues in software productivity and sustainability,
including software engineering for CSE and tools for software project
management.
For example, code repositories that support version control are increasingly used as
a management tool for projects of all sizes.  Teaching students at
all levels to routinely use version control will increase their
productivity and allow them to participate in open-source software
projects in addition to better preparing them for many jobs in
large-scale CSE.

Researchers in CSE fields (and from
governments, funding agencies, and the public) also have experienced growing concern about the lack of
reproducibility of many scientific results based on code and data that is
not publicly available and often not properly archived
in a manner that allows future confirmation of the results.
Many agencies and journals are beginning to require open sharing of data and/or
code.  CSE education should include training in the techniques
that support this trend, including data management and provenance,
licensing of code and data,
full specification of models and algorithms within publications,
and archiving of code and data in repositories
that issue permanent identifiers such as DOIs.

The important issues of ethics and privacy also come into play with an increased community focus on sharing data and codes. These topics are an essential part of CSE education---they should be covered as part of a class as well as reinforced through research process and mentoring.

\subsection{Changing Educational Infrastructure}

As we think about CSE educational programs, we must also consider the changing external context of education, particularly with regard to the advent of digital educational technologies and their associated impact on the delivery of education programs.

One clear impact is an increased presence of online digital materials, including digital textbooks, open educational resources, and massive open online courses (MOOCs). Recent years have already seen the development of online digital CSE resources, as well as widespread availability of material in fields relevant to CSE, such as HPC, machine learning, and mathematical methods.
An opportunity exists to make better community use of current materials, as well as to create new materials. There is also an opportunity to leverage other resources, such as Computational Science Graduate Fellowship essay contest winners\footnote{\url{https://www.krellinst.org/csgf/outreach/cyse-contest}} and archived SIAM plenaries and other high-profile lectures.
The time is right for establishing a SIAM working group that creates and curates a central repository linking to CSE digital materials and coordinates community development of new CSE online modules.
This effort could also be coordinated with an effort to pursue opportunities in continuing education.

Digital educational technologies are also having an impact on the way university courses are structured and offered. For example, many universities are taking advantage of digital technologies and blended learning models to create ``flipped classrooms," where students watch video lectures or read interactive online lecture notes individually and then spend their face-to-face class time engaged in active learning activities and problem solving. Digital technologies are also offering opportunities to unbundle a traditional educational model---introducing more flexibility and more modularity to degree structures. Many of these opportunities are well suited for tackling the challenges of building educational programs for the highly interdisciplinary field of CSE.

%% file: outcomes-education.tex

\begin{table}

\caption{Learning outcomes desired of a student graduating from a CSE
  Ph.D. program.  Italicized text denotes differences in learning
  outcomes for programs with a broadly applicable CSE focus (left) and
  a domain-driven focus in a particular field of science or
  engineering (right). Learning outcomes that are common to both types
  of Ph.D. programs span left and right columns.\label{tab:LO}}
\begin{center}
\begin{tabular}{|p{2.9in}|p{2.9in}|}\hline
\textbf{CSE Ph.D. with Broadly Applicable CSE Focus}
& \textbf{CSE Ph.D. with Domain-Driven Focus in Field X}\\ \hline

Combine mathematical modeling, physical principles, and data to derive,
analyze, and assess {\em models across a range of systems (e.g.,
statistical mechanics, continuum mechanics, quantum mechanics,
molecular biology)}.
& Combine mathematical modeling, physical principles and data to
derive, analyze, and assess a {\em range of models within field X}. \\ \hline

\multicolumn{2}{|p{5.8in}|}
{Demonstrate graduate-level depth in devising, analyzing, and evaluating new methods and algorithms for
computational solution of mathematical models (including parallel,
discrete, numerical, statistical approaches, and mathematical analysis).} \\ \hline

Demonstrate {\em mastery} in code development to exploit parallel and
distributed computing architectures and other emerging modes of computation in
algorithm implementation.
& Demonstrate {\em proficiency} in code development to exploit parallel and
distributed computing architectures and other emerging modes of computation in
algorithm implementation. \\ \hline

\multicolumn{2}{|p{5.8in}|}
{Be aware of available tools and techniques from software engineering, their strengths and their weaknesses; select and apply techniques and tools from software engineering to build robust, reliable, and maintainable software.} \\ \hline

\multicolumn{2}{|p{5.8in}|}
{Develop, select, and use tools and methods to represent and visualize
computational results.} \\ \hline

\multicolumn{2}{|p{5.8in}|}
{Critically analyze and evaluate results using mathematical and data
analysis, physical reasoning, and algorithm analysis, and understand
the implications on models, algorithms, and implementations.} \\ \hline

\multicolumn{2}{|p{5.8in}|}
{Identify the sources of errors in a CSE simulation (such as modeling
errors, code bugs, premature termination of solvers, discretization
errors, roundoff errors, numerical instabilities), and understand how to diagnose them and work
to reduce or eliminate them.} \\ \hline

Appreciate and explain the context of decision-making as the end use
of many CSE simulations, and as appropriate be able to formulate,
analyze, and solve CSE problems in control, design, optimization, or
inverse problems.
& Appreciate and explain the context of decision-making as the end use
of many CSE simulations, and as appropriate be able to formulate,
analyze and solve CSE problems in control, design, optimization or
inverse problems {\em as relevant to field X}. \\ \hline

\multicolumn{2}{|p{5.8in}|}
{Understand data as a core asset in computational research, and
demonstrate appropriate proficiencies in processing, managing, mining,
and analyzing data throughout the CSE/simulation loop.} \\ \hline

\multicolumn{2}{|p{5.8in}|}
{Demonstrate the ability to develop, use, and analyze sophisticated
computational algorithms in data science and engineering, and
understand data science and engineering as a novel field of
application of CSE.} \\ \hline

Demonstrate graduate-level {\em proficiency in one domain in science or engineering}.
& Demonstrate graduate-level {\em depth in domain knowledge in field X}. \\ \hline

\multicolumn{2}{|p{5.8in}|}
{Communicate across disciplines and collaborate in a team.} \\ \hline \hline
\end{tabular}
\end{center}
\end{table}

%% file: conclusions.tex
\bigskip
\section{Conclusions and Recommendations}
\label{sec:conclusions}
\pagebudget{1}
\team{All}

\subsection{Summary}
Over the past two decades, computational science and engineering has become tremendously successful and influential at driving progress and innovation in the sciences and technology.
CSE is intrinsically interdisciplinary, and as such it often suffers from the entrapments created by disciplinary boundaries.
While CSE and its paradigm of quantitative computational analysis and discovery are permeating increasingly many areas of science, engineering, and beyond, CSE has been most successful when realized as a clearly articulated focus within its own well-defined academic structures and its own targeted funding programs and aided by its own focused educational programs.
The past decade has seen a comprehensive broadening of the application fields and methodologies of CSE. For example, mathematics-based computing is an important factor in the quantitative revolution that is sweeping through the life sciences and medicine, and powerful new methods for uncertainty quantification are being developed that build on advanced statistical techniques.

Quantitative and computational thinking is becoming ever more important in almost all areas of scientific endeavor.  Hence, CSE skills and expertise must be included in curricula across the sciences, including the biomedical and social sciences.
A well-balanced system of educational offerings is the basis for shaping the future of CSE. 
Creating a unique identity for CSE education is essential.
Dedicated CSE programs have been created up to now only in a relatively small number of universities, mostly in the United States and Europe. More such undergraduate and graduate-level (masters and Ph.D.) programs in CSE are necessary in order to train and prepare the future generation of 
CSE scientists to make new scientific and engineering discoveries. This core CSE education will require designing dedicated curricula; and where such programs already exist, continuous adaptation is needed to address the rapidly changing landscape of CSE.

\bigskip
\subsection{Central Findings}
\begin{itemize}

\item \textbf{F1:}
\textbf{CSE as a discipline.}
CSE has matured to be a discipline in its own right.
It has its own unique research agenda, namely, to invent, analyze, and implement
broadly applicable computational methods and algorithms
that drive progress in science, engineering, and technology.
A major current focus of CSE is to create truly predictive capability in science.
Such CSE-based scientific predictions will
increasingly become the foundation of
technical, economic, societal, and public policy advances and decisions
in the coming decades.

\item \textbf{F2:}
\textbf{Algorithms and software as research artifacts.}
Innovations in mathematical methods and computational algorithms lie at the core of CSE advances.
Scientific software, which 
codifies and organizes algorithmic models of reality,
is the primary means of encapsulating CSE research
to enable advances in scientific and engineering understanding.
CSE algorithms and software can be created, understood, and properly employed
only by using a unique synergy of knowledge that combines an
understanding of mathematics, computer science, and target problem areas.

\item \textbf{F3:}
\textbf{CSE and the data revolution.}
CSE methods and techniques are essential in order to capitalize
on the rapidly growing ubiquitous availability of scientific and technological data,
which is a major challenge that calls for the development of new numerical methods.
In order to achieve deeper scientific benefit, data analytics must proceed beyond the exposition of correlations.
CSE develops new statistical computing techniques that are efficient at scale,
and it incorporates physical models informed by first principles to
extract from the data insights that go far beyond what can be recovered
by statistical modeling alone.

\end{itemize}

\subsection{General Recommendations}

\begin{itemize}
\item \textbf{R1:}
Universities and research institutions should {\bf expand CSE to
realize its broad potential for driving scientific and technological
progress} in the 21st century. This requires removing disciplinary
boundaries, engaging with new application areas, and developing new
methodologies.  Multidisciplinary research and education
structures where CSE is a clearly articulated focus should be
increasingly encouraged.

\item \textbf{R2:}
Funding agencies should \textbf{develop focused and sustained funding programs} that address the specific needs of research in CSE. These programs should acknowledge the multidisciplinary nature of CSE and account for specific research agendas of CSE, including CSE algorithms and software ecosystems as critical instruments of a novel kind of predictive science and access to leading high-performance computing facilities.

\end{itemize}

\subsection{Recommendations for CSE Education}

\begin{itemize}
\item \textbf{E1:}
Universities should \textbf{strengthen and broaden computational thinking in all relevant academic areas and on all levels}. This effort is vital for driving scientific, technological, and societal progress and needs to be addressed systematically at the university level as a crucial factor in workforce development for the 21st century.

\item \textbf{E2:}
\textbf{Dedicated CSE programs at all university degree levels should be created} to educate future core CSE researchers for jobs in the private and government sectors, in research laboratories, and in academia. New CSE-centric teaching materials are required to support such programs.

\item \textbf{E3:}
The \textbf{common core of CSE and data science}, as well as their synergy, \textbf{should be exploited} in educational programs that will prepare the computational and data scientists of the future. Aided by scientific visualization and interactive computational experiments, CSE is a powerful motivator for study in the STEM disciplines at pre-university levels. Outreach materials are required.

\end{itemize}

\newpage

%% file: acknowledgments.tex
\section*{Acknowledgments}
\label{sec:acknowledgments}

This report is an outcome of a workshop in August 2014 on {\em Future
Directions in CSE Education and Research}, sponsored by the Society
for Industrial and Applied Mathematics~\cite{siam:homepage} and
the European Exascale Software Initiative~\cite{EESI-generic}.
We gratefully acknowledge La Joyce Clark and Cheryl Zidel of Argonne
National Laboratory for workshop support.
We also thank Jenny Radeck of Technische Universit\"{a}t M\"{u}nchen for
assistance with graphics and Gail Pieper of Argonne National Laboratory for
editing this document.

The first author gratefully acknowledges support by the Institute of Mathematical Sciences
at the National University of Singapore during the work on this report.
The work of the third author was supported by the U.S. Department of
Energy, Office of Science, under contract number DE-AC02-06CH11357.